# DYNAMISM IN THE SOLAR CORE

Attila Grandpierre, Ph. D.
Konkoly Observatory
H-1525 Budapest, P. O. Box 67, Hungary
grandp@iif.hu

**Abstract.** Recent results of a mixed shell model heated asymmetrically by transient increases in nuclear burning indicate the transient generation of small 'hot spots' inside the Sun somewhere between 0.1 and 0.2 solar radii. These hot bubbles are followed by a nonlinear differential equation system with finite amplitude non-homologous perturbations which is solved in a solar model. Our results show the possibility of a direct connection between the dynamic phenomena of the solar core and the atmospheric activity. Namely, an initial heating about $DQ_0 \approx 10^{31}$-$10^{37}$ ergs can be enough for a bubble to reach the outer convective zone. Our calculations show that a hot bubble can arrive into subphotospheric regions with $DQ_{final} \approx 10^{28}$ - $10^{34}$ ergs with a high speed, up to 10 km s$^{-1}$, approaching the local sound speed. We point out that the developing sonic boom transforms the shock front into accelerated particle beam injected upwards into the top of loop carried out by the hot bubble above its forefront traveling from the solar interior. As a result, a new perspective arises to explain flare energetics. We show that the particle beams generated by energetic deep-origin hot bubbles in the subphotospheric layers have masses, energies, and chemical compositions in the observed range of solar chromospheric and coronal flares. It is shown how the emergence of a hot bubble into subphotospheric regions offers a natural mechanism that can generate both the eruption leading to the flare and the observed coronal magnetic topology for reconnection. We show a list of long-standing problems of solar physics that our model explains. We present some predictions for observations, some of which are planned to be realized in the near future.

## 1. INTRODUCTION

Recently, a series of new results (Ehrlich, 2007; Wolff, 2009; Hiremath, 2009) indicated that the dynamism of the solar core, which we suggested in our previous papers (Grandpierre, 1996, Paper I.; Grandpierre and Agoston, 2005; Paper II) can be of interest.

A suggestion made in our Paper I was carried out in more details, and the results led to a new mechanism to explain the Ice Ages of the Earth (Ehrlich, 2007). Moreover, Wolff (2009) pointed out that the g-modes indicate the presence of 'hot spots' we predicted in Papers I and II. He had shown that a whole list of fundamental problems of solar physics have the prospect of being understood if a mixed shell below 0.2 $R_\odot$ permits the excitation of *g*-modes that form rigidly rotating sets. The overlapping *g*-modes can represent 3% of the solar luminosity. A mixed shell model heated asymmetrically by transient increases in nuclear burning indicates the transient generation of small 'hot spots' inside the Sun somewhere between 0.1 and 0.2 solar radii. Moreover, Hiremath (2009) had pointed out that the observed quasi-periodicities of solar activity indices in the range of 1-5 years are explained due to perturbations of the strong toroidal field structure and, variation of very long period solar cycle and activity phenomena such as the Maunder and grand minima is explained to be due to coupling of long period poloidal and toroidal MHD oscillations. Since the origin and nature of the solar activity cycle is still a formidable and unsolved problem, it seems to be timely to revise the widespread notion that the solar core is simple.

Not so long ago, Bahcall (1989, 43) characterized the general view as follows: "The Sun's interior is believed to be in a quiescent state and therefore the relevant physics is simple". But there is a not yet suitably acknowledged reason to think that the solar core has a definite dynamic nature; namely, that the solar core is a magnetized plasma. This fundamental fact escaped attention due to another popular myth considering the Sun as being merely a "luminous ball of gas" (Ridpath, 1997, 450). But it is easy to see that the Sun is not a "ball of gas", since it consists largely from ionized particles, and it is penetrated by a magnetic field (as it is shown below). Therefore, the Sun is a vast plasma system coupled to time varying rotation and activity. The Sun can be regarded as a laboratory in which all the four fundamental interactions: electromagnetism, gravitation, weak and strong interactions, are coupled to each other in vast dimensions. The fact that the solar interior is in a plasma state has a much far reaching significance than it is generally realized nowadays. The electromagnetic interaction is 39 orders of magnitude stronger than the gravitational one; correspondingly, it is enormously richer in nonlinear interactions. Therefore, even if the charges largely balance each other, the remaining small unbalanced forces may dominate behavior. This is the reason why plasma systems show collective behavior. Plasma systems show an enormous variety of dynamic phenomena: being highly electrically conductive, they respond to magnetic fields. Magnetic fields show a tendency to instabilities, have a complex and time varying spatial structure, generating various time varying current systems, filaments, sheets, and jets, manifesting extremely rich behavior. Goossens (2003, 1) emphasized: Plasmas are extremely complicated systems fundamentally different from classic neutral gases, especially when there is a magnetic field present. We will show below by quantitative estimations that magnetic fields play an important role in the plasma of the solar core.

In contrast to a simplified, fictitious plasma of some theoretical approaches, Alfven (1968) had shown that real plasma show much more complicated behavior. (1) Quite generally a magnetized plasma exhibits a large number of *instabilities*. (2) A plasma has a tendency to produce *electrostatic double layers* in which there are strong localized electric fields. Such layers may be stable, but often they produce oscillations. (3) If a current flows through an electrostatic double layer (which is often produced by the current itself), *the layer may cut off the current*. This means that the voltage over the double layer may reach any value necessary to break the circuit (in the laboratory, say $10^5$ or $10^6$ V; in the magnetosphere, $10^4$-$10^5$ V; in solar flares, even $10^{10}$ V). The plasma "explodes", and a high-vacuum region is produced. (4) Currents parallel to a magnetic field (or still more in absence of magnetic fields) have a tendency to pinch; i.e., to *concentrate into filaments* and not flow homogeneously. The inevitable conclusion from phenomena (1) through (4) above is that *homogeneous models are often inapplicable*. Nature does not always have a *horror vacui* but sometimes a *horror homogeneitatis*. In contrast to gases, a plasma, particularly a fully ionized magnetized plasma, is a medium with basically different properties. Typically it is strongly inhomogeneous and consists of a network of filaments produced by line currents and surfaces of discontinuity. These are sometimes due to current sheaths and, sometimes, to electrostatic double layers (Alfven & Arrhenius, 1976, Chapter 15).

Regarding the electric aspect of the problem, we can realize that electric instabilities, e.g. plasma microinstabilities generally lead to development of current filaments. In general, the plasma can support electric currents (Goossens, 2003, 1). As a rule, the development of instability is accompanied by an increase in the electric field strength, which can attain large values. Consequently, even in the absence of intense external fields, relatively strong fields can still occur spontaneously in plasma due to the growth of instability (Tsytovich, 1970, 1). In a highly conductive plasma electric instabilities, especially in changing magnetic fields, develop through complex, nonlinear effects in a way that necessarily lead to a strong amplification of electric currents. Highly amplified electric currents involve extremely strong local heating in the simultaneous presence of inhomogeneous electromagnetic fields and high densities. Recently, Chang et al. (2003) pointed out that the basic MHD equations admit fluctuations to develop that can generate fluctuation-induced nonlinear instabilities reconfiguring the topologies of the magnetic fields.

Apparently, one of the most important keys to the dynamic processes of the solar core lies in the magnetic field. Recently, Gough and McIntyre (1998) had shown that the maintenance of vertical and horizontal shears characterizing the tachocline require their confinement by an underlining magnetic field having a strength of ~1 G just beneath the tachocline. They have argued also that a nonzero interior poloidal field $B_i$ is necessary to explain the observed closely uniform rotation of the radiative zone, and estimated that $B_i \approx 10^3$ G well below the top of the radiative zone if the magnetic field $B_i$ deep in the radiative interior is the remnant of a primordial field. Friedland and Gruzinov (2004) had shown by solar model calculations that there are many modes of the toroidal field with lifetimes long enough to survive until today. Therefore, the toroidal field in the radiative zone of the Sun can, in principle, have complex structure. The strength of toroidal fields, being entirely confined to the radiative zone, is not subject of the above bound. Friedland and Gruzinov (2004) gave their lowest upper bound to be $B_{FG}$~$2.1*10^6$ G.

It was Hiremath (2005a) who determined the poloidal and toroidal parts of the magnetic field profiles in the radiative zone as a function of distance from the solar centre, following the calculations of Hiremath and Gokhale (1995) who used the information of angular velocity as inferred from helioseismology and solved self-consistently the axisymmetric and incompressible MHD equations. Hiremath (2005a) found toroidal field strengths up to $10^4$ G.

These numerical values for the strength of the magnetic field allow us to estimate the active role of magnetic field in amplifying the dynamism of the solar core's plasma. The criterion for neglecting magnetic effects in the treatment of a problem in gas dynamics is that the Lundquist parameter $L_u = (4\pi)^{1/2}\sigma B l_c c^{-2}\rho^{-1/2}$ (measuring the ratio of the magnetic diffusion time to the Alfven travel time), where $\sigma$ is the electric conductivity in e.s.u., $B$ is the strength of the magnetic field in Gauss, $l_c$ is a characteristic length of the plasma in centimeter, $\rho$ is the mass density in gcm$^{-3}$, and $c$ is the speed of light), is much less than unity, $L_u \ll 1$ (Alfven & Arrhenius, 1976, Chapter 15). Now for the solar core $\sigma \approx 10^{17}$ e.s.u., $B \approx 2*10^{-3}$ to $2*10^{6}$ G, $l_c \approx 10^{10}$ cm, $\rho \approx 10^2$ g cm$^{-3}$, and so $L_u \approx 7*10^{2-11}$, therefore $L_u \gg 1$. This means that plasma effects may play a dominant role in the dynamism of the solar core. For the toroidal field strength of the solar core obtained by Hiremath (2005a) $B_H \approx 10^4$ G, we obtain $L_u \approx 3*10^9 \gg 1$; for Friedland's and Gruzinov's (2004) lowest upper bound $B_{FG} \sim 2.1*10^6$ G, the result is $L_u \approx 7*10^{11} \gg 1$.

Gervino, Lavagno, and Quarati (2001) determined that the plasma parameter $\Gamma (=e^2/R_D kT$; here 'e' is the charge of the electron, $R_D$ is the Debye screening length, 'k' the Boltzmann constant, 'T' is the temperature of the plasma; $\Gamma$ measures the mean Coulomb energy potential to the thermal kinetic energy in the solar interior is $\Gamma \approx 0.1$, therefore long-range many-body interactions and memory effects play a significant role. Moreover, they pointed out that the plasma frequency, having a value of 3 to $6*10^{17}$ sec$^{-1}$, is of the same order like the collisional frequency; and the screening radius is of the order of the interparticle distance. In addition to many-body collisional effects, electric microfields are present, modifying the usual Boltzmann kinetics. These estimations indicate that highly nonlinear and complex plasma effects may actually and directly play a dominant role in the dynamism of the solar core.

Instead of being a fast rotator as expected, recent helioseismic observations had shown that the solar core rotates almost rigidly. This means that the spin down of the solar core proceeds continuously even nowadays. Hiremath (2001) had shown that differential rotation at the base of the convective zone is more likely than uniform rotation. Therefore it is plausible to allow a small rate of radial differential rotation in the solar core, e.g. $w_1 \sim 10^{-3} w_0$ of the rate present in the convective zone. Such a differential rotation generates dissipative processes and, certainly, magnetic instabilities. Certainly, there have to be a coupling between the solar radiative interior and the convective zone, especially if a magnetic field connects these regions. In the case of the Earth, the core-mantle coupling is actually an important trigger contributing to earthquake occurrence (Wang et al., 2004). Therefore, one can consider it plausible that sporadic, singular, localized energy liberation processes, similar to earthquakes, also occur in the solar core. Now Gough and McIntyre (1998) as well as Hiremath (2005a) pointed out that the coupling poloidal field has strength in the magnitude of $B_p \approx 1$ G, a definite value enough to look after measurable consequences. This result makes it plausible to assume that magnetic coupling between the solar core and outer regions generates local energy release in the solar core.

## 2. ESTIMATIONS OF LOCAL HEATING: SPATIAL AND TEMPORAL SCALES

Now let us estimate some of the possible effects of magnetic instabilities present in the radiative core of the Sun. Hiremath's equation (Hiremath, 2005a, eq. 4) offers an estimation for the maximum size of the magnetic instabilities L. Based on simple dimensional analysis and assuming that meridional velocity is very small compared to either the poloidal ($B_p$) and toroidal ($B_t$) field strengths or angular velocity, and assuming that both the poloidal and toroidal parts have large diffusion time scales, one can get a relation between differential rotation term $w_1$ and those parameters ($B_p$, $B_t$, $\eta$, L) as follows:

$w_1 \sim (\eta/L^2)*(B_t/B_p)$,

where $\eta$ is the magnetic diffusivity and L is the length scale. By taking the values computed by Hiremath (2005a) of $B_p$ and $B_t$ in the radiative zone and for a weak differential rotation rate in the solar core $w_1 \approx 10^{-3} w_0$, instabilities may not affect the length scales larger than 100 km. Since the magnetic field structure which Hiremath (2005a) proposed is a large-scale one, we don't expect any instability on larger scales. At the same time, these estimations had shown that one has to admit that on the length scales of L < 100 km in the radiative zone, plasma instabilities may exist.

The timescale of the simple magnetohydrodynamic Tayler instability (Tayler, 1973, Spruit, 2002) is very short, of the order of hours and days (Tayler, 1973, Goossens, Tayler 1980). It was shown that the most unstable perturbations have a very small wavelength (Goossens, Veugelen, 1978). Zirin (1988, 48) argued that the rate of growth of magnetic instabilities is given by $\tau_{magn} \approx L/v_A$, where *L* is the characteristic spatial scale, and '$v_A$' is the Alfven velocity $v_A = B/(4\pi\rho)^{1/2}$. For example, with $L=10^7$ cm and $B_H=10^4$ G one obtains $v_A=5*10^4$ cm s$^{-1}$, and so $\tau_{magn}=200$ s.

Let us estimate the heating energy available by magnetic reconnection. With $w_1 \approx 10^{-3} w_0$ and $B_H=10^4$ G, when taking a linear size for a spherical region $L=10^7$ cm, one obtains for the local heating a value around $\Delta Q_0(magn) \approx 2*10^{28}$ ergs; and with $B_{FG}=2.1*10^6$ G, $L=10^7$ cm, $\Delta Q_0(magn) \approx 7*10^{32}$ ergs. The calculations of Grandpierre and Ágoston (2005) had shown that already a heating $\Delta Q_0 \approx 10^{27}$ ergs can be enough to generate a buoyant force which drive the heated region upwards so that the so-formed bubble can make a distance more than its linear size. Depending on the concentration of the heating energy, our calculations indicated that $\Delta Q_0 \approx 10^{32}$ ergs can be enough to drive the bubble upwards to make a significant portion of the solar radius (Figs. 1 and 2, Grandpierre and Ágoston, 2005). We note that we consider a plasma system in a strong magnetic field. Magnetic energy dissipation may be expected to occur in a filamentary, highly concentrated form, around narrow current channels. Therefore we can think that reconnection favors highly localized regions to heat.

With a density $\rho \approx 10^2$ g cm$^{-3}$ characteristic to the inner solar core, this latter amount of heating energy may lead in the volume with a linear scale $L=10^7$ cm, to a heating $\Delta T \approx 10$ K. It is easy to see that when $\Delta Q_0 \approx 10^{27}$ ergs heat a smaller, e.g. $L \approx 10^5$ cm region, the arising relative temperature surplus will be $\Delta T \approx 10^7$ K, a value large enough leading to the formation of a hot bubble traveling a path much larger than its diameter. Without any concentration of dissipation, for a certain relative temperature surplus, e.g. when $(\Delta T)_0/T_S \approx 0.1$ and $R \approx 10^6$ cm, we will need an initial heating $\Delta Q_0 \approx 10^{35}$ ergs. More generally, for $\rho \approx 1$ to $10^2$ g cm$^{-3}$, $R \approx 10^5$ cm to $10^7$ cm, we will need $\Delta Q_0 \approx 10^{32}$ ergs to $10^{38}$ ergs for $(\Delta T)_0/T_S \approx 0.1$ that is probably necessary for successful surfacing of the bubble. In Table 1, we present the range of relative heating available by magnetic reconnection.

We emphasize that the basic point of our consideration is that there have to exist some kind of a singular heated region of the solar core. We point out that the lawful development of a singular hot bubble within the solar core has a principal significance. If a certain temperature surplus is present locally, nonlinear couplings within the extremely complex plasma conditions present in the solar core will certainly lead, from time to time, in dependence on local conditions, to amplification of this perturbation to values that are higher by orders of magnitude. Therefore, the estimations presented above merely illustrate situations typically developing in certain localized regions of the solar core – and the numerical calculations below will determine the conditions within which the perturbations will be amplified to observable consequences.

Besides the energies dissipated in magnetic instabilities, rotational energy dissipation is also indicated to become concentrated into small heated regions. In an inhomogeneous body penetrated by a magnetic field, rotational energy dissipation is generally manifested in intermittent local events. During the last $4.6*10^9$ years the solar core has been spun down from a 50 times higher value at the zero-age main sequence (Charbonneau and MacGregor, 1992) $E_{rot,0} \approx 10^{45}$ ergs to the present one $E_{rot,present} \approx 2.4*10^{42}$ ergs (Allen, 1963, 161). From Fig. 2a of Charbonneau and MacGregor (1992) one can read that the present rate of solar spin-down corresponds to $(\Delta E_{rot}/\Delta t)_{present} \approx 2*10^{34}$ ergs year$^{-1}$. It is a general view that the dissipation of rotational energy is used to drive the dynamo, and, in general, solar activity in the solar envelope. We note that the main part of the rotational energy is dissipated in the spin-down of the solar core. Now since inhomogeneous electromagnetic field is indicated in the solar core, it seems plausible to assume that $\Delta E_{rot}$ is dissipated intermittently and highly localized in the solar core, in a form suitable to drive activity phenomena. We can obtain estimation for the possible amount of heating on the basis that seven "hot spots" was observed during five solar cycles (Bai, 2003), therefore the average rate of formation of hot spots is cca. 0.1 year$^{-1}$. If we identify the formation of a hot spot with a rotational dissipation event, we obtain that the rotational heating has a magnitude $\Delta Q_0(rot) \approx 2*10^{35}$ ergs.

The presence of local heated regions in the solar core is also suggested by Burgess et al. (2003). They indicated the presence of density fluctuations in the deep solar core as a result of a resonant process similar to coronal heating, and had shown that the energy that is transferred from the helioseismic gravity modes into magnetic Alfven modes with density fluctuations leads to strong local heating. They pointed out that the measured spectrum of helioseismic waves does not rule out density variations with amplitudes as large as 10% on scales close to L~100 km (again, the same spatial scale limit as that of magnetic instabilities). It is easy to show that such a density variation involves a dissipated heating energy $\Delta Q_0$(resonant heating) $\approx 10^{40}$ ergs. Now let us see a few timely words about the gravity modes, the necessary input to Burgess's model.

Recently, Garcia et al. (2004) pointed out that some patterns are detected by Turck-Chieze et al. (2004) using GOLF data during the last solar minimum that can be interpreted in terms of gravity modes. The rotation profile is now clearly established down to the limit of the core. In order to progress toward the core and reduce the uncertainties in the radiative region, gravity modes should be measured. Today we cannot arrive at a firmly established conclusion, but we cannot exclude the possible detection of several components of gravity mode candidates. The research of gravity modes detection in the solar core will continue. Turck-Chieze et al. (2004) are confident of making conclusions on the existence of gravity modes in the observations before the end of the life of SOHO in 2007. The possibility that gravity modes will become detectable in the near future, and they may show traces of dynamism, makes it an urgent task to consider theories and models predicting the dynamism of the solar core. In this paper, we developed detailed numerical calculations showing that dynamism of the solar core proceeds in a yet overlooked manner: through singular, individual events like a hot bubble. Stability considerations of the solar core were made only for the shells of the solar core.

Paterno, Rapisarda and DiMauro (1997) reconsidered nonradial thermal instabilities in the solar core for internal, infinitesimal, homologous, i.e., shellular perturbations. They have found that the solar core is stable against such perturbations. Remarkably, already on the basis of their result we may conjecture that the solar core is close to instability for finite amplitude nonradial perturbations. This circumstance is due to the fact that the heating timescale they obtained – for homologous perturbations produced by nuclear heating: $\tau_{growth} \sim 4 \times 10^6$ years is only slightly higher than the cooling one, arising from radiative diffusion: $t_{decay} \sim 7 \times 10^5$ years (see their Table 2; at the solar centre). In this paper, we point out that perfect spherical symmetry is impossible in the real Sun. The consequences of significant singular deviations from spherical symmetry seem to be overlooked in the context of instabilities in the solar core. For a highly localized, singular heating, stability analysis was not performed yet, and carrying out such an investigation may be regarded as an important task. We note here that the necessity of a dynamic solar core model is already indicated by many independent theoretical and observational arguments (Grandpierre, 1990, 1996), and a trend towards the dynamical representation of the stars is noted (Turck-Chieze, 2001).

Besides the above theoretical arguments underpinning the sporadic localization of energy liberation in the plasma of the solar core, we also may have some observational support indicating the presence of heated regions and flare-like phenomena in the solar radiative interior. There are well-founded reasons telling that the observations of Toutain, Kosovichev (2001) and Chaplin et al. (2003) may be signs of flare-like events in the deep solar core. Chaplin et al. (2003) found an anomalous event at late March 1998 supplying additional energy to solar activity and low-l solar p-modes. This event raised the velocity power ($V^2$, which is directly proportional to the total energy of a mode) by 22% above the zero change level; the predicted value for this epoch in the cycle, however, is of the order of $\sim (-5)$ %. By our best knowledge, similar energy enhancements of p-modes are observed until now only in relation to flares (Haber et al., 1988; Kosovichev & Zharkova, 1998). Chaplin et al. (2003) noted that the increase of energy supply is coincident in time with the southern hemisphere onset of cycle 23, with a major emission of particles and the appearance of major surface activity on this hemisphere. Remarkably, Benevolenskaya (1999) had shown that the transition from cycle 22 to cycle 23 clustered in the very same fixed longitudinal regions. Recognizing that such activity enhancements are usually related to active regions with especially high flare activity, and that this event occurred well below the photosphere, one may assume that the increase of energy supply is related to a certain localized event somewhere in the solar interior. Because this event is energetic and localized, one may apply the term "flare-like event". Now it is an interesting point raised by Bai (2003) – see below - that the hot spots had to form in regions independent of toroidal magnetic flux tubes, in the radiative zone. Therefore, the formation of hot spots should be related to flare-like events are indicated to occur in the radiative zone by the following observations.

Bai (2002) paid attention to the fact that solar flares from the southern hemisphere during cycle 23 are found to be concentrated in a pair of hot spots rotating with a synodic period of 28.2 days, slightly surpassing the range of rotational periods observed both on the surface and in the convective zone in the latitude zone extending from -35 degrees to 35 degrees, 26 to 28 days. Moreover, Bai (2003) has been found that the hot spots of this double hot-spot system are separated by about 180 degrees in longitude. Many hot-spot systems last for more than one solar cycle, and therefore the mechanism(s) generating them must be independent of the dynamo working in and around the convective zone. Since the toroidal fields are found around the top of the radiative zone, the mechanism(s) generating the hot spots must act below the zone containing the toroidal flux tubes. Taking into account the facts those hot-spot systems set up frequently in a 180 degrees longitudinal separation, and that they have an anomalous rotation rate from 25 to 29 days, one may seem plausible to find the origin of hot spots deep in the solar core. Actually, helioseismic measurements allow such anomalously rotating layers or regions if their spatial scales are less than 100 km. Therefore, the localization of the source of hot spots suggests the presence of localized hot regions deep in the solar core. It seems plausible to allow that the source of hot spots may be related to the heating events which produce the increased energy supply for solar activity and p-modes in March 1998 (Chaplin et al, 2003).

These theoretical and observational results all indicate that the solar core tend to form sporadic localized heated regions. Therefore, it is important to consider the development of finite amplitude local heating in the solar core. In this paper, we show that in the solar radiative interior sporadically generated hot bubbles may travel significant distances towards the surface. We found that the generation of heated regions presents a new, yet not considered type of instability that lends certain dynamism to the solar core which may have a fundamental significance in the origin of solar activity.

In concluding this section, we present a small table (Table 1.), summarizing our findings on the amplitudes of local heating, arising from the given mechanisms. While below $T \sim 10^8$ K cooling has two important mechanisms, volume expansion and radiative diffusion, above $T \sim 10^8$ K radiant cooling will be dominant, and therefore our approach is not valid already. Therefore, in Table 1, we indicated large heating simply by $\Delta T_0/T_S > 10$.

Table 1. Available amplitude of heating and initial energy surplus with the different heating mechanisms

| Heating mechanisms | Amplitude of heating | Initial temperature surplus |
| $\Delta Q_0$(magnetic) | $2*10^{28}$ - $7*10^{32}$ ergs | $\Delta T_0$(magn)/$T_S$~$10^{-11}$ to >10 |
| $\Delta Q_0$(rotational) | $2*10^{35}$ ergs | $\Delta T_0$(rot)/$T_S$~$10^{-3}$ to >10 |
| $\Delta Q_0$(resonant heating) | $10^{40}$ ergs | $\Delta T_0$(res)/$T_S$ > 10 |

These results are confirmed by our detailed numerical calculations presented below. The large values of temperature surplus of heated bubbles illustrate that all the three mechanisms that are indicated to be plausible sources of heating can lead to significant heating. Therefore, it is instructive to carry out numerical simulations, taking into account the most important heating and cooling processes.

The investigation of bubble-like perturbations creates a new situation in comparison to the already considered shellular case. Bubble formation can couple hydrodynamic instabilities to thermal perturbations. Therefore, it is interesting to follow such finite amplitude bubble-like perturbations individually by numerical computations. In this way, we can determine the parameters of the arising hydrodynamic movements, including the distance a heated bubble may travel, and this parameter may be an important indicator of the dynamism of the solar core.

## 3. ON THE REAL PHYSICS OF THE SUN AND THE BASIC EQUATIONS OF HOT BUBBLES

We pointed out in the Introduction that the plasma nature of the solar core presents a complication that has a far reaching significance. We may add that solar activity is a macroscopic phenomenon that occurs as a collective phenomenon involving the cooperation of enormous numbers of particles. The problem of how macroscopic phenomena arise from properties of the microscopic constituents of matter is basically a quantum mechanical one (Sewell, 1986, 4). The Sun is composed of approximately $10^{56}$ interacting particles of several species. At a microscopic level, therefore, its properties are governed by the Schrödinger equation for this assembly of particles. However, the Schrödinger equation of the Sun is extremely complicated: indeed its extreme complexity represents an essential part of the physical situation. Due to this extreme complexity, collective quantum fields become dominating over the individual fields of particles.

Actually, the Sun is a system that is extremely more complicated than its constituents. Being much more than the sum of its constituents, the Sun belongs to the most complex systems of the universe, showing an unusually wide range of emergent complex phenomena: solar activity. In the case of the Sun, not only the collective quantum fields become dominating over the individual ones, but also the gravitational and the electromagnetic fields. Moreover, these dominating cooperative collective fields do interact with each other as well as with the nuclear fields and energy production. Therefore, the Sun is extremely more complex than complex systems in the Earth. This extreme complication leads to computational problems. In order to make the problem solvable, one has to find the physically most interesting aspects of this complexity that may offer a simplification leading to a tractable formulation of the problem.

The preliminary consideration of the Introduction served calling attention to the problem of highly localized, singular heating in the solar core.

If we restrict ourselves to describe the rise of the bubbles, the extremely complicated problem simplifies to a tractable one that can be described by the Navier-Stokes equation as the equation of motion, the energy equation and the equation of state.

$$\rho(\partial v_i/\partial t + v_k \partial v_i/\partial x_k) = f_i - \partial p/\partial x_i + \eta \partial^2 v_i/\partial x_k \partial v_k \quad (i=1, 2, 3) \quad (1)$$

where $f_i$ is the i-component of the total force acting per unit mass and p is the pressure. The conservation of energy tells that

$$dU/dt + p/\rho \, \text{div} \, v = \varepsilon_N - 1/\rho \, \text{div} \, (F_R + F_c) + (p/\rho)\partial v_k/\partial x^i , \quad (2)$$

where $U$ is the total thermal energy, $\varepsilon_N$ is the liberated nuclear energy per unit mass and time, $F_R$ and $F_c$ are the radiative and conductive fluxes.

The equation of state is

$$p = (R_g/\mu) \rho T \quad (3)$$

where $\mu$ is the dimensionless mean molecular weight, and $R_g$ is the gas constant.

## 4. BASIC ESTIMATIONS FOR THE CASE OF A HEATED BUBBLE

We formulate the following scenario: a dissipation event heats a local parcel of matter in the solar interior at some depth (at first, we selected $r=0.1 \, R_{Sun}$). We calculated how this initial perturbation generates a heated bubble (or heated region, in the absence of bubble formation) which is already in pressure equilibrium with its surroundings.

A heated bubble is not in hydrostatic equilibrium with its surroundings. In the first approximation of the Navier-Stokes equation, the motion of a heated bubble is determined by the equality of the buoyant ($F_b = Vg \, \Delta\rho$) and frictional ($F_f = K/2 \, v^2 \, S \, \rho$) forces, where $S$ is the cross section of the bubble, $\rho$ is the density of the bubble, $V$ is its volume, $K$ is the coefficient of turbulent viscosity, $\rho_S$ and $\Delta\rho$ are the density of the surroundings and the density difference between the bubble and its surroundings, and $g$ is the gravitational acceleration. Equating these forces, $v^2 \, (K/2) \, (S/V) = g \, \Delta\rho/\rho_S$. Now assuming pressure equilibrium between the bubble (referred with no index) and its surroundings (referred with index S), $\rho T = \rho_S T_S$, $\Delta \rho/\rho_S = (1 - T_S/T)$. Taking $K = 1$ (Öpik, 1958), we obtain for the bubble's velocity $v = (8/3 R_g (1 - T_S/T))^{1/2}$. With local heating events in the solar core $\Delta T/T_S \approx 0.1$ (from our Table 1) yields $T_S/T \approx 8/9$, $g \approx 2*10^5$ cm s$^{-2}$, R=$10^5$-$10^6$ cm, the rising speed of the bubble is v$\approx$2 to 7 *$10^5$ cm s$^{-1}$.

Now we turn to the energy equation. The heated bubble is not in thermal equilibrium with its surroundings. Below $T \approx 10^8$ K the radiation energy and pressure may be neglected compared to the material energy and pressure. The radiation energy must, of course, not be neglected in the flux term. In a co-moving frame, without energy sources, when radiation is the most effective dissipative factor, the energy equation may be simplified to the form

$$\partial U/\partial t = -1/\rho \, \text{div} \, F_R = -1/\rho \, \text{div} \, (D_R \, \text{grad} \, E_R), \quad (4)$$

where $E_R = aT^4$ is the radiation energy density, and $a$ is the radiation-density constant. Assuming that one can apply the diffusion approximation $D_R = 1/3 \, c \, l_{ph}$, where $c$ is the speed of light, and $l_{ph} = 1/(\kappa\rho)$ is the mean free path of a photon, $\kappa$ is a mean absorption coefficient. In the case of spherical symmetry, the corresponding diffusive radiative flux is

$$F_R = -(4ac/3\kappa\rho) T^3 \partial T/\partial r. \qquad (5)$$

Now returning to the simplified energy equation (5), with $U=C_p T$, and integrating it to the whole volume of the bubble,

$$C_p (\partial T/\partial t)V = - (1/\rho) 4\pi R^2 F_R. \qquad (6)$$

From this equation the thermal adjustment time is estimated, in a linear approximation, writing for $\partial T/\partial t \sim -\Delta T/\tau_{adj}$, and for $\partial T/\partial r \sim \Delta T/R$, following Kippenhahn and Weigert (1990, 44) as:

$$\tau_{adj} = \kappa\rho^2 C_p R^2/(16\sigma T^3), \qquad (7)$$

where $\sigma=ac/4$ is the Stefan-Boltzmann constant ($\sigma= 5.67*10^{-5}$ erg cm$^{-2}$K$^{-4}$s$^{-1}$). With typical values, $\kappa = 2$ cm$^2$ g$^{-1}$, $\rho = 90$ g cm$^{-3}$, $C_p=2.1 *10^8$ erg K$^{-1}$ mole$^{-1}$, $T = 10^8$ K, $R = 10^6$ cm, $\tau_{adj} = 3*10^3$ s, while for $T = 10^7$ K, $\tau_{adj} = 4*10^6$ s.

To obtain a preliminary picture on the question whether heated bubbles may travel a distance larger than their characteristic sizes, first we determined the relevant timescales of this process. It is a favorable method because it offers a fast and easy way to obtain a first view on the relations between heating and bubble rise.

In order to put this fundamental thermal timescale into the context we are interested in, we define a time-scale for the rise of the bubbles as $\tau_{rise} = l_T/v$, where $l_T$ is the temperature scale height (in the solar core at r=$R_{Sun}$/10, $l_T \sim 1.5*10^{10}$ cm). With v=1.5 * $10^5$ to 1.5 * $10^6$ cm s$^{-1}$, $\tau_{rise} \sim 10^5$ to $10^4$ s, respectively. This means that for (at least) moderate heating (when $T/T_S>1.0001$) the bubble may move so fast that its thermal cooling is slower than the decrease of the temperature of its environment on its path rising towards the surface (for a moment, we ignore the cooling arising from volume expansion; more detailed results are given later on). In such a case, the bubble cannot adjust its temperature to its environment, and the heating may lead to the formation of a bubble and its self-maintaining rise upwards - even if we disregard from any internal energy source.

The timescale of cooling of the bubbles arising from adiabatic volume expansion may be calculated following Gorbatsky (1964). Starting from

$$Q=C_V m T = 2 \pi R^3 p,$$
$$(dQ/dt)_{exp} = -p \, d(4/3\pi R^3)/dt$$
$$2\pi p (3 R^2) (dR/dt)_{exp} + 2 \pi R^3 (dp/dt)_{adiab} = -p \, 4 \pi R^2 (dR/dt)_{exp}$$
$$(dR/dt)_{exp} = -R/5 (1/p(dp/dt))$$
$$\tau_{exp} = -(1/5 (1/p (dp/dr)) v)^{-1} = 5 H_p/v.$$

Around r=0.1$R_{Sun}$, the pressure scale height $H_p = |1/p(dp/dr)|^{-1} \sim 7.3 *10^9$ cm. For R=$10^5$-$10^6$ cm and with v $\sim 10^3$ - $10^6$ cm s$^{-1}$, $\tau_{exp}$ is usually in the range of $3*10^7$ - $10^4$ s. This is an important result, since it indicates that the rise time $\tau_{rise}\sim10^4 – 10^5$ s and the adiabatic expansion timescale $\tau_{exp}\sim10^4$-$3*10^7$ s somewhat overlap. More concretely, as Table 2 indicates, the rise time is lower than the expansion timescale; therefore the bubble may make significant distances, since it rises faster than it cools by adiabatic expansion. Actually, with v$\sim10^6$ cm s$^{-1}$, to make a distance $R_{Sun}\sim7*10^{10}$ cm, the bubble needs $\tau_{final}\sim10^5$ s.

It is easy to make the calculations given in Tables 2 and 3. We obtained that $\tau_{exp} \approx 5\ H_p/v$, $\tau_{diff} = \kappa \rho^2 C_P R^2/(16\sigma(T^3-T_S^3))$, $\tau_{cool} = (\tau_{exp}^{-1} + \tau_{diff}^{-1})^{-1}$, $\tau_{rise} \approx 1.5*10^{10}/v$. We used $\tau_{diff}$ instead of $\tau_{adj}$ since we took into account the fact that when the heated bubble temperature approaches the temperature of its surroundings, its diffusive radiative flux decreases to zero. We tested these preliminary estimations with detailed numerical calculations, considering that the material heated by the heat wave of radiative diffusion expanding from the bubble is coupled to it (see Gorbatsky, 1964).

TABLE 2
The time-scales with $R_0 = 3*10^6$ cm and different $T_0$. The bubble rises from $r=0.1R_{Sun}$. The case with $\tau_{diff} = \kappa \rho \rho_S C_P R^2/[16\sigma(T^3-T_S^3)]$.

| $T_0$ [in $10^7$ K] | $\tau_{exp}$ | $\tau_{diff}$ | $\tau_{cool}$ | $\tau_{nucl}$ | $\tau_{rise}$ |
|---|---|---|---|---|---|
| 1.74 | $5.7*10^4$ | $7.8*10^6$ | $5.6*10^4$ | $1.1*10^{14}$ | $2.3*10^4$ |
| 5.0 | $3.3*10^4$ | $6.6*10^4$ | $2.2*10^4$ | $6.9*10^7$ | $1.3*10^4$ |
| 9.0 | $3.0*10^4$ | $6.2*10^3$ | $5.1*10^3$ | $2.6*10^5$ | $1.2*10^4$ |
| 20.0 | $2.9*10^4$ | $2.5*10^2$ | $2.5*10^2$ | $7.8*10^1$ | $1.2*10^4$ |

TABLE 3.
The time-scales when the bubble rises from $r=0.65R_{Sun}$ (a) with $\tau_{diff} = \kappa \rho \rho_S C_P R^2/[16\sigma(T^3-T_S^3)]$, with $R_0 = 5*10^5$ cm and (b) $\tau_{diff} = \kappa \rho_S^2 C_P R^2/[16\sigma(T^3-T_S^3)]$, with $R_0 = 2*10^6$ cm.

| $T_0$ [in $10^7$ K] | $\tau_{exp}$ | $\tau_{diff}$ | $\tau_{cool}$ | $\tau_{nucl}$ | $\tau_{rise}$ |
|---|---|---|---|---|---|
| 6.2 (a) | $1.4*10^5$ | $2.1*10^2$ | $2.1*10^2$ | $3.5*10^{10}$ | $8.7*10^4$ |
| 6.2 (b) | $2.4*10^5$ | $3.6*10^{-1}$ | $3.6*10^{-1}$ | $3.5*10^{10}$ | $4.4*10^4$ |

The local enhancement of nuclear energy liberation needs a heating timescale $\tau_{nucl} = C_p T/\varepsilon \nu$ (Grandpierre, 1990) to be shorter than the timescale of the cooling processes $\tau_{exp} = 5\ H_p/v$, $\tau_{adj} = \kappa \rho^2 C_p R^2/(16\sigma T^3)$, where $C_p$ is the specific heat at constant pressure, T the temperature of the heated region, $\varepsilon$ is the rate of energy liberation by nuclear reactions, $\nu$ is the exponent in the $\varepsilon \sim T^\nu$ relation, $H_p$ is the pressure scale height, 'v' the velocity of the heated region, $\kappa$ is a mean absorption coefficient, $\rho$ and R is the density and the radius of the heated region, $\sigma$ is the Stefan-Boltzmann constant ($\sigma = 5.67 * 10^{-5}$ erg cm$^{-2}$ K$^{-4}$s$^{-1}$). With typical values ($\kappa \sim 2$ cm$^2$ g$^{-1}$, $\rho \sim 100$ g cm$^{-3}$, $C_p \sim 3 * 10^8$ erg K$^{-1}$ mole$^{-1}$, $T \sim 10^7$ K, $R \sim 10^6$ cm, $H_p \sim 7*10^9$ cm, $v \sim 10^3$-$10^6$ cm/s), $\tau_{adj} \sim 7 * 10^6$ s, $\tau_{exp} \sim 3*10^4$-$10^7$ s, $\tau_{nucl} \sim 10^{16}$ s, while for $T \sim 10^8$ K, $\tau_{adj} \sim 7*10^3$ s, $\tau_{exp} \sim 3*10^4$ - $10^7$ s, $\tau_{nucl} \sim 1$ s.

This means that when the sporadic and localized energy dissipation processes indicated in our Table 2 heat a small macroscopic region more then tenfold, above $10^8$ K, nuclear energy liberation may make the region explosive, since the diffusion of radiation and volume expansion together cannot cool the heated volume effectively on such a short timescale. At such temperatures, the cooling timescales are more than three orders of magnitude larger than that of the nuclear heating; therefore, local thermonuclear runaway will develop. The volume of the heated region may explosively increase until it forms such a heated bubble that, accelerated by the buoyant force, rises up and transports outwards most of the produced surplus energy.

From these estimations we can recognize the remarkable situation that all the four relevant time-scales determining the behavior of the bubbles $\tau_{rise}$, $\tau_{exp}$, $\tau_{adj}$ and $\tau_{nucl}$ are comparable to each other. Therefore, it is important to consider the case by more detailed numerical calculations and determine if there exist suitable conditions for triggering instability.

## 5. METHOD OF CALCULATION

We start by picking up a certain determined virtual value for the radius of the bubble $R_V$ and for its initial virtual temperature surplus $n=T_V/T_S$. These virtual values are not physical values but they soon will turn into realistic values by pressure equilibration. At the very first phase of the bubble formation the density of the heated bubble is $\rho_V=\rho_S$, and $T_V=nT_S$, $Q_V=2\pi\mu_0^{-1}R_g R_V^3\rho_V T_V$, $p_V=p_0=np_S$, $n>1$. Then we determine the parameters of the bubble which underwent pressure equilibration and is already in pressure equilibrium with its environment (denoted with indices "0"), $\rho_0=\rho_V n^{-3/5}$, $R_0=R_V n^{-1/5}$, $Q_0=Q_V n^{-2/5}$, $T_0=Q_0(\mu_0/2\pi)(R_0^3 R_g\rho_0)^{-1}$, $m=\rho_0(4\pi/3)R^3$. At $t=0$, $m_+(t=0)=0$ (here 'm' and '$m_+$' are the initial mass of the heated bubble, and the mass of the volume heated by radiative diffusion of the bubble, respectively). Then we pick up a certain set of time steps, and determine the values of the parameters in the next time step.

We worked with a fourth order Runge-Kutta method to solve the differential equation system. Our calculations differ from such previous ones like Rosenbluth and Bahcall (1973), and Paterno, Rapisarda and DiMauro (1997), who worked with $\Delta\rho = 0$, and $\Delta(T\mu)=0$, since they considered merely homologous, strictly non-radial perturbations. In our calculations, we allowed non-homologous, singular perturbations, non-vanishing only in a highly localized region, without a strict local hydrostatic equilibrium, and so the heated region may have initially pressure surplus, too. After a transient period lasting for a few seconds pressure equilibrium sets up, and the bubble is hotter and less dense than its surroundings, $\Delta p \sim \Delta(T\rho) = 0$, but $\Delta T_0 \neq 0$ and $\Delta\rho_0 \neq 0$.

We solved the differential equation system with a numerical code (a simpler version of the code is described in Grandpierre, Ágoston, 2005). We neglected the radiation pressure in all the terms except the diffusive one. Our method may be regarded as working well for our purposes below $10^8$ K, since the estimated error in each quantity is smaller than 15 %.

## 6. RESULTS AND DISCUSSION

The calculated timescales show us that the characteristic rise time of the bubbles are comparable (or shorter) than their combined cooling time scales, therefore the bubbles are able to rise significant distances in the radiative core. We note that our calculations involved turbulent drag only approximately. It is possible that in reality turbulence may disturb the hot bubbles in such a rate that they become disintegrated before running distances much larger than their diameter. At the same time, the formation of enveloping plasmoid around the bubbles may be favorable for enhancing their lifetimes. Bubbles in water usually survive rise distances much larger than their linear sizes. Hot bubbles enveloped in a plasmoid may also survive large distances.

The obtained results lead to an important conjecture: namely, there exist a yet unexplored type of stellar instability within the solar core and similar stellar radiative interiors. Fig. 1 illustrates the effect of density difference. The difference between curves 1 and 1', 2 and 2', 3 and 3' corresponds to a change from $\tau_{diff} = \kappa \rho \rho_S C_P R^2/(16 \sigma (T^3-T_S^3))$ to $\tau_{diff} = \kappa \rho_S^2 C_P R^2/(16 \sigma (T^3-T_S^3))$. The mechanism(s) compressing the bubble on its path may be due to the aerodynamic drag and, more importantly, to the tension of environmental deeply rooted magnetic field lines that the bubbles met on their pathway and that the rising bubbles elongate and push at their forefront upwards. Magnetic fields when forming plasmoid structure around the bubble may serve simultaneously as an accelerator agent through the accompanying magnetic buoyancy effect. When the rising bubble does not decrease its density so fast, due to plasmoid confinement, radiation will escape from its surface in a lesser rate, and so the bubble can travel larger distances, as our detailed calculations show.

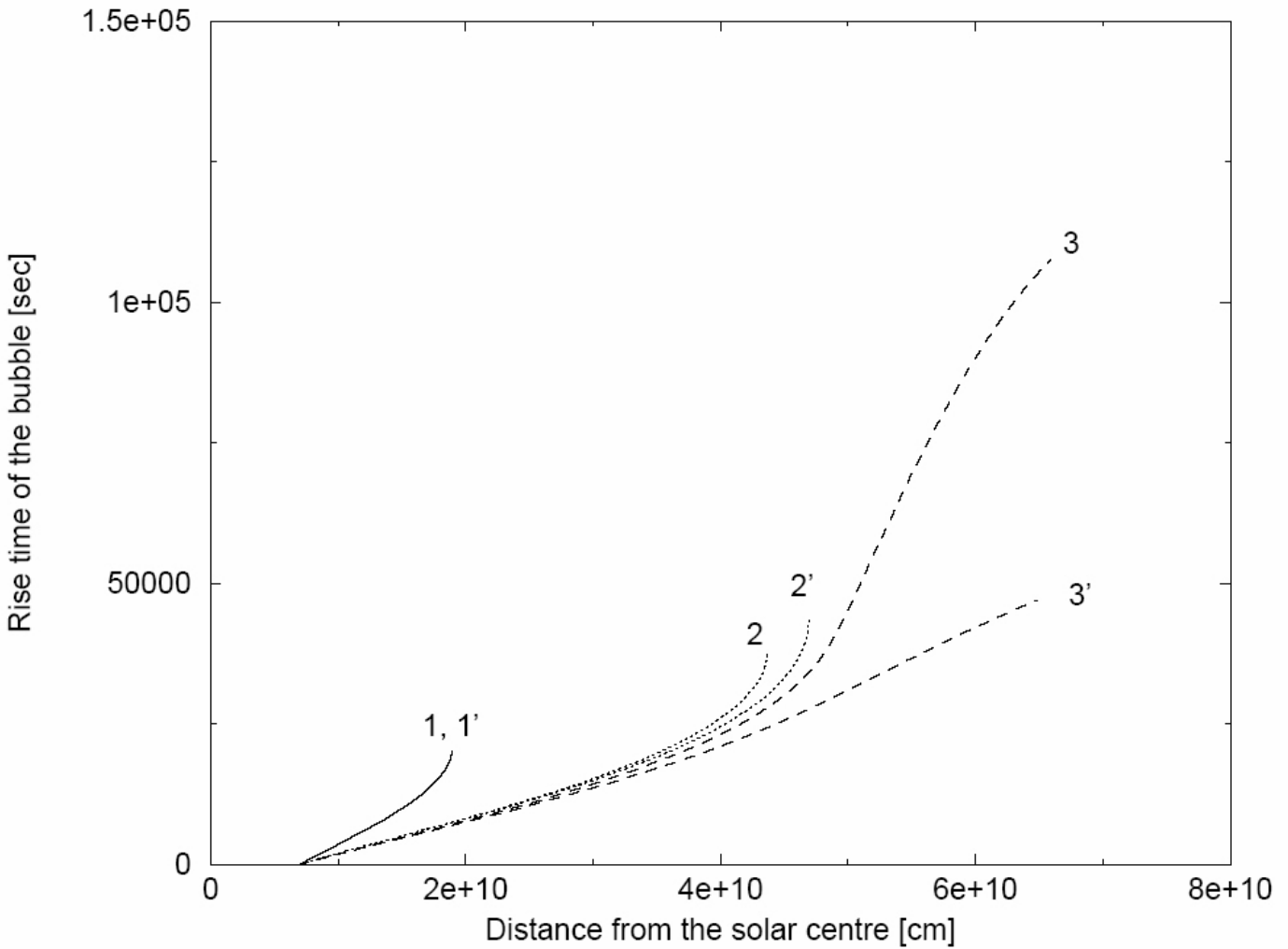

Fig. 1 illustrates the effect of density difference. The difference between curves 1 and 1', 2 and 2', 3 and 3' corresponds to a change from $\tau_{diff} = \kappa \rho \rho_S C_P R^2/(16 \sigma (T^3 - T_S^3))$ to $\tau_{diff} = \kappa \rho_S^2 C_P R^2/(16 \sigma (T^3 - T_S^3))$.

We found that when heating above a certain energy threshold is present, it can directly initiate from time to time large-amplitude individual motions of heated bubbles that can travel significant distances within the solar body. Our calculations indicated that this threshold is around $\Delta Q_0 \approx 10^{27}$ ergs. Larger bubbles and larger heating may lead to bubbles traveling much larger distances. Fig. 2 shows the effect of the initial spatial size of the bubble. Indices 1, 2, 3 and 4 refers to $R_0 = 10^6$ cm, $2 * 10^6$ cm, $3 * 10^6$ cm, and $4 * 10^6$ cm, respectively. Although smaller bubbles make also significant distances, there exist a certain critical range of spatial sizes (corresponding to different amount of initial heating, see Table 1), above which the bubble may reach the surface regions.

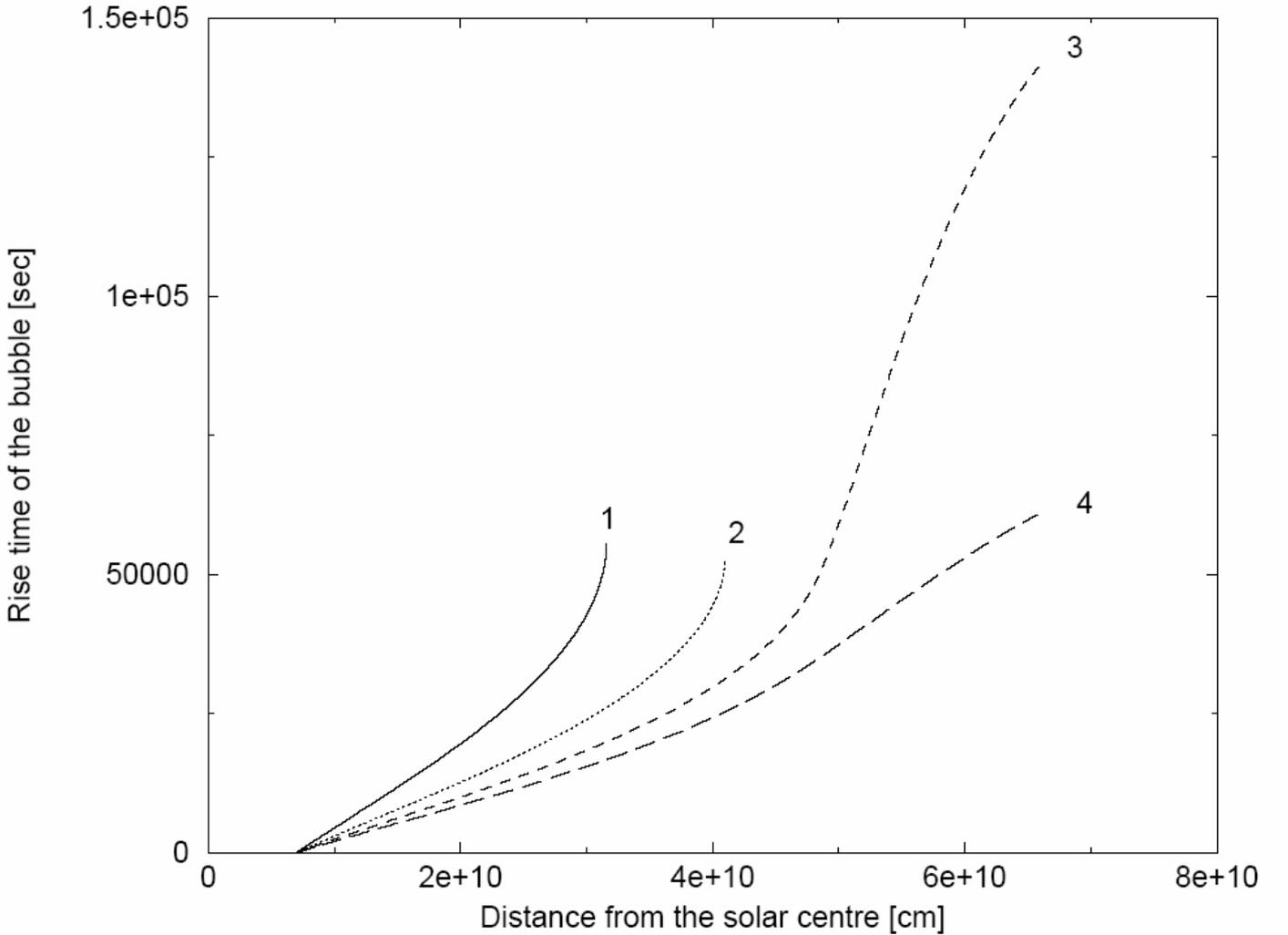

Fig. 2 shows the effect of the initial spatial size of the bubble. Indices 1, 2, 3 and 4 refers to $R_0 = 10^6$ cm, $2 * 10^6$ cm, $3 * 10^6$ cm, and $4 * 10^6$ cm, respectively.

Fig. 3 shows the evolution of the rising speed of the bubble. The bump around r=1.2 to 1.4*10$^{10}$ cm in the rising speed 'v' is related to the slow increase of radius $R$ and the local maximum of the gravitational acceleration $g_{max}$ = 2.4 * 10$^5$ cm s$^{-2}$ at r ~ 1.05*10$^{10}$ cm that is followed by a fast decrease of $g$ since 'v' is proportional to $(Rg)^{1/2}$. Fig. 3 shows that for T<10$^8$ K, from r=0.1 $R_{Sun}$ the bubble may need 10$^{37}$-10$^{38}$ ergs to reach the surface, if we neglect the possibly favorable effect of involvement into a buoyant magnetic plasmoid here.

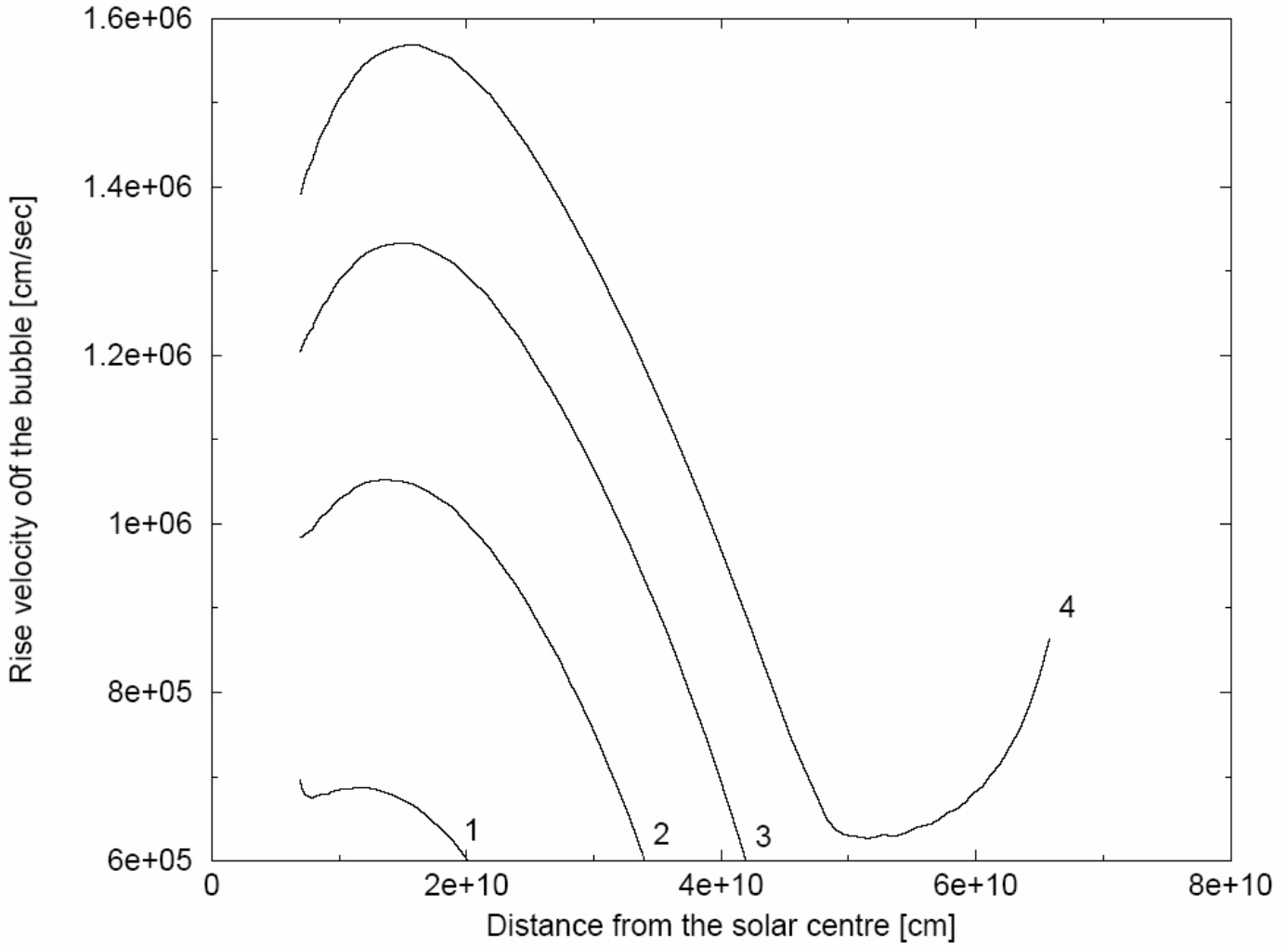

Fig. 3 shows the evolution of the rising speed of the bubble.

Fig. 4 shows the rate of hydrogen deficiency of the bubbles relative to their local environments in dependence of the distance from the solar centre. Bubbles move upwards, therefore they represent the chemical composition of deeper regions, and so their surfacing may be related to local chemical abundance anomalies. We will discuss the possible observational consequences of heavy element enhancements in solar flares shortly below.

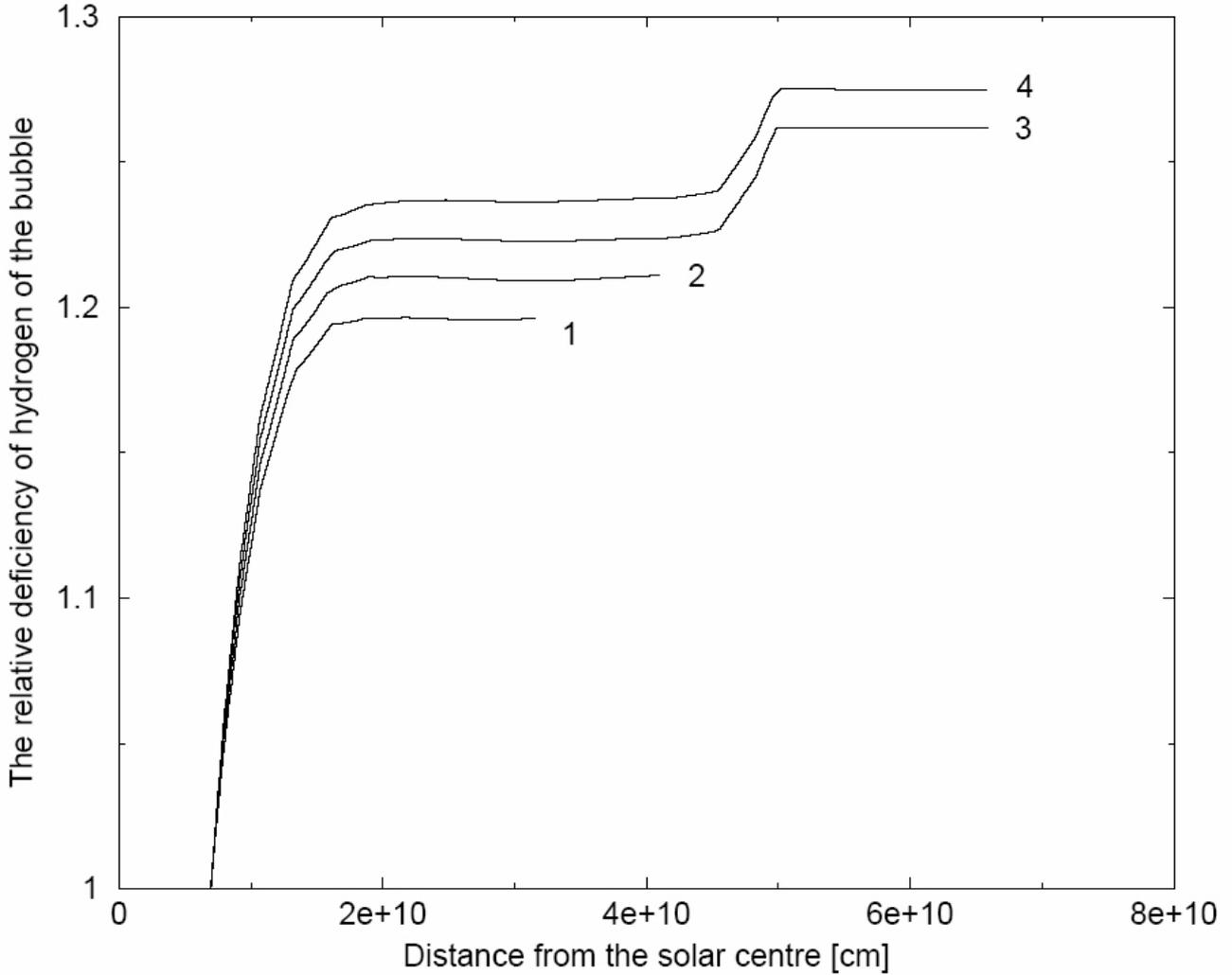

Fig. 4 shows the rate of hydrogen deficiency of the bubbles relative to their local environments in dependence of the distance from the solar centre.

Fig. 5 shows the rise of the bubble for the case when the bubble starts to rise from $r=0.1$ $R_{Sun}$. One may notice that the increase of the initial temperature surplus of the bubble is not automatically helpful for the surfacing of the bubble. On the contrary, our calculations had shown that overly high heating causes faster cooling when the 'velocity' of radiative diffusion $v_{diff} = R/\tau_{diff}$ becomes higher than $v(=v_{rise})$, the bubble will cease to rise at lower distance from the solar centre. The stopping of the $R_0 = 4*10^5$ cm bubble for $T_0 = 5*10^7$ K will occur at $r_{final} = 5.07*10^{10}$ cm, and for $T_0 = 5*10^6$ K sooner, at $r_{final} = 4.85*10^{10}$ cm. This "cutting effect" of radiative diffusion may be effective in constraining the surfacing of the bubbles to a narrow range of heating, especially when the bubbles are formed not far below from the tachocline. Nevertheless, at deeper regions the cutting effect of radiative diffusion is less effective, but here other limiting factors may be effective, like heating energy input constraints. We note here that we obtained for the smallest energy surplus of the surfacing bubble a value of $\Delta Q_0 = 3.9*10^{31}$ ergs for $T_0 = 1.5*10^7$ K, when the bubble is formed and started its rise from $r_0=0.65$ $R_{Sun}$. We note that we did not find bubbles reaching the convective zone for $R_0<4*10^5$ cm. This fact also means that $\Delta Q_0 > 3*10^{31}$ ergs is necessary for the bubbles to reach the subsurface regions, if we ignore here the possible role of being enveloped into a buoyant magnetic structure as a result of plasma interactions generating the bubble.

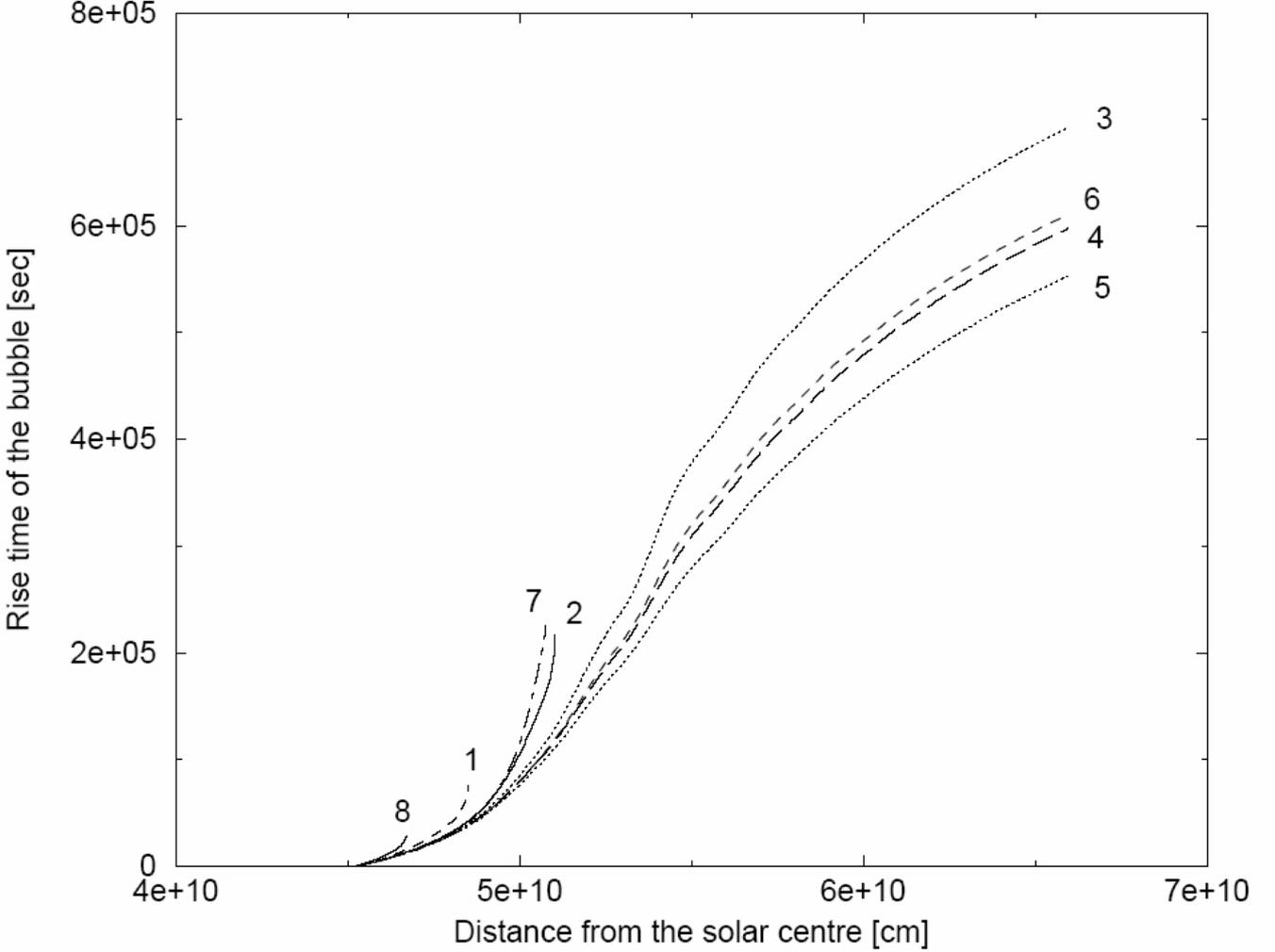

Fig. 5 shows the rise of the bubble for the case when the bubble starts to rise from r=0.1 $R_{Sun}$. The stopping of the $R_0 = 4*10^5$ cm bubble for $T_0 = 5*10^7$ K will occur at $r_{final} = 5.07*10^{10}$ cm, and for $T_0 = 5*10^6$ K sooner, at $r_{final} = 4.85*10^{10}$ cm.

In Fig. 6 we plotted the rise of the bubbles for three cases with $\tau_{diff}$ proportional to $\rho_S^2$, starting from $0.1R_{Sun}$, from $0.4R_{Sun}$ and from $0.65R_{Sun}$. The common characteristics of these bubbles are that they are marginally able to reach the convective zone (and, therefore, the subphotospheric regions). Their path is close to each other near the bottom of the convective zone at $r=4.9*10^{10}$ cm. The bubble rising from $0.1R_{Sun}$ have an initial energy surplus $\Delta Q_0 \sim 9.3*10^{37}$ ergs (dotted curve), the other one rising from $0.4R_{Sun}$ corresponds to $\Delta Q_0 \sim 4*10^{35}$ ergs (solid curve), the third one from $0.65R_{Sun}$ (dashed curve) to $\Delta Q_0 \sim 4*10^{31}$ ergs. The dashed curve runs to $\tau_{rise} \sim 6*10^5$ s at $r=0.98R_{Sun}$. Bubbles surfacing from $0.65R_{Sun}$ have rising times $(5$ to $6)*10^5$ s, almost independently from their initial temperature surplus.

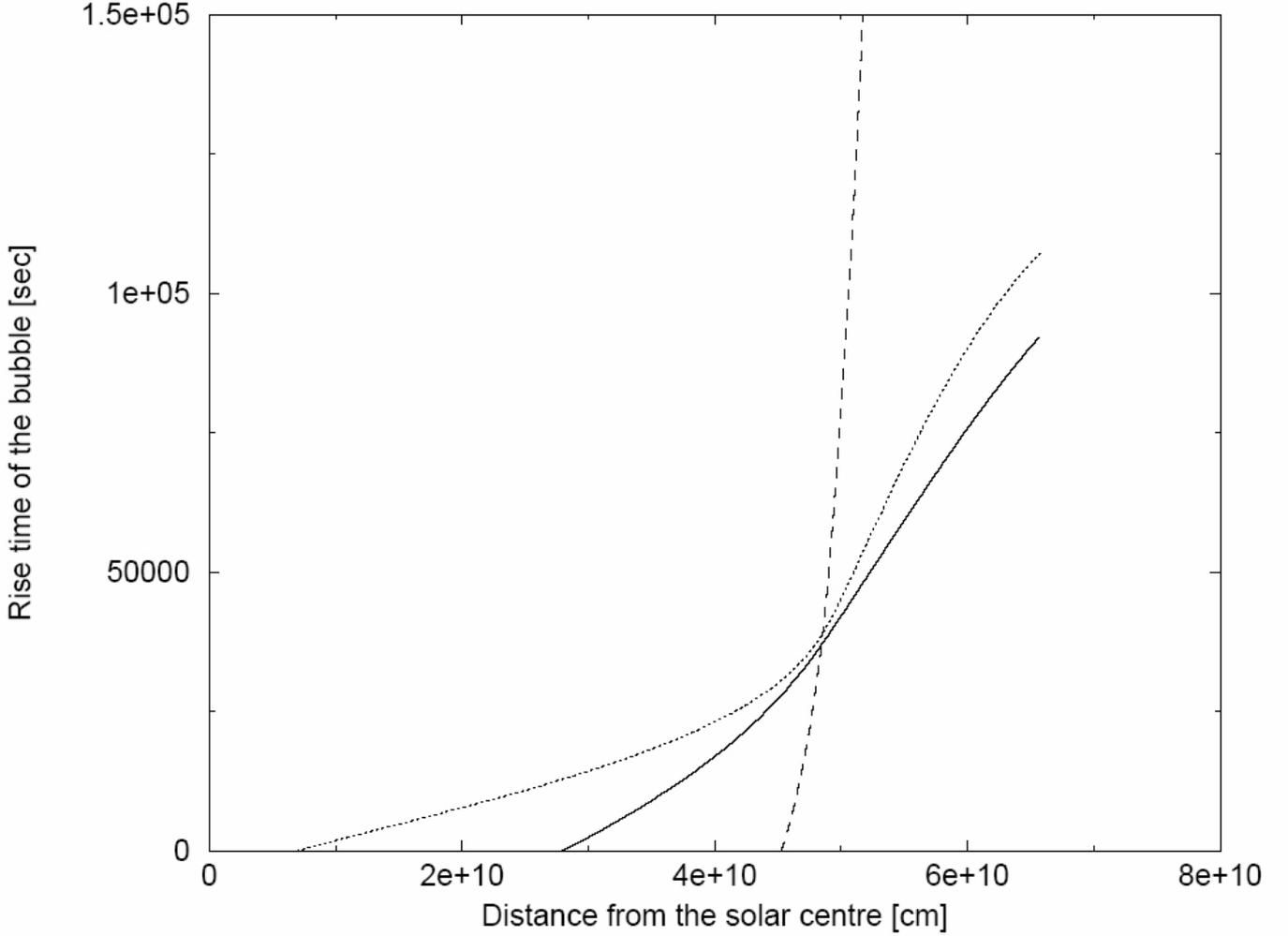

Fig. 6 shows the rise of the bubbles for three cases with $\tau_{diff}$ proportional to $\rho_S^2$, starting from $0.1R_{Sun}$, from $0.4R_{Sun}$ and from $0.65R_{Sun}$.

We give some quantitative results regarding the arrival of hot bubbles into the subphotospheric regions in Table 4.

Table 4.
The relations between subphotospheric bubble energies, bubble masses, final relative temperature surplus, final velocity and final radius

| $\Delta Q_0$[ergs] | $\Delta Q_{final}$[ergs] | $m_{final}$[g] | $T_{final}/T_S$ | $v_{final}$[km s$^{-1}$] | $R_f$[km] |
|---|---|---|---|---|---|
| $4.2*10^{31}$ | $1.32*10^{30}$ | $1.6*10^{18}$ | 1.01 | 0.6 | 35 |
| $4.0*10^{35}$ | $2.34*10^{33}$ | $3.8*10^{20}$ | 1.10 | 3.9 | 209 |
| $1.1*10^{38}$ | $2.6*10^{34}$ | $1.4*10^{22}$ | 1.04 | 4.3 | 680 |

We note that in Table 4 we presented only marginal cases, seeking the minimum initial energy surplus necessary to reach subphotospheric regions. Therefore, for larger initial energies, the hot bubbles may reach the near surface regions with higher speed and energy surplus.

Remarkably, the obtained energies, masses and sizes are in an apparently noticeable fit with the observed range of the same quantities characteristic to solar flares. This is the more interesting since the problem of flare energetics – as we will show here - has been never solved satisfactorily. It is many times emphasized that it is a belief that magnetic sources may supply the flare with enough energy. Priest and Forbes (2002, 317) express their belief that "The energy source [of solar flares] must be magnetic since all the other possible sources are completely inadequate". But Craig and McClymont (1999, 1045) stated that "the ability of any reconnection mechanism to explain the massive, explosive magnetic collapse of the solar flares remains questionable". Holman (2003) noted: It is generally believed that the flare energy is derived from the coronal magnetic field. However, we have not been able to establish the specific energy release mechanism(s) or relative partitioning of the released energy between heating, particle acceleration and mass motions. Machado (2001) wrote: "The most intriguing aspect of impulsive phase physics resides in the mechanism that leads to the release of, say, $10^{30}$ ergs in $10^2$ s…The resolution of the impulsive phase enigmas will be addressed through a worldwide coordinated program of flare observations to start in the year 2000." Yet it seems that the worldwide coordinated program did not bring the long-awaited solution: "The energization process [of solar flares] is still unknown" (Masuda, 2004). "Magnetic reconnection is generally believed to play a crucial role in solar coronal activity. A central paradox is that magnetic reconnection must occur at very small scales in order to be fast enough but must be directly affect the largest scales in order to matter" (Longcope, 2005). This means that the resolution of impulsive phase enigmas still waits to be clarified.

The widespread conviction claiming that flare energies have to be supplied by magnetic reconnection is based on the notions that i.) Magnetic energies may supply enough energy to flares, and, ii.) There is no any real alternative mechanism that may supply energy even comparably well. At first, we show that magnetic energies, in contrast to widespread beliefs, seem to be overly poor energy sources for solar flares. Secondly, we show that there is a powerful alternative energy source to energize flares.

When an engineer constructs the plan of a bridge, he must have plan the bridge to survive the possible largest challenges. Similarly, it is not the small, or moderate, but the large, and, first of all, largest flares that have a crucial importance regarding flare theories. Solar activity involves a wide range of phenomena on many spatial and temporal scales, corresponding to a wide variety of energy flows. In the case of a small flare, more energy source may be important, than in the case of largest flares, where the most effective energy source is indicated to be dominant. It is clear that we need to know about the most effective possible energy source of flares. Benz (2001) starts his review paper "Solar Flare Observations" with the introducing sentence: "Flares are caused by the release of magnetic energy up to some $10^{27}$ J (some $10^{34}$ ergs) in the solar atmosphere within a few minutes". This means that $E_f > 10^{34}$ ergs and $t_{imp} \sim 10^2$ (<$10^3$) s. Li, Mickey and LaBonte (2005) found that the energy released by their white light flare is $\sim 10^{33}$ ergs – their flare is classified in X-rays as an X3-class flare. There are some dozens of flares of higher X-classes, one of the highest estimated is the 04/11/2003 being estimated as X28. Recently, it became clear that giant flares may be a feature of solar cycle, since giant flares are produced in both of the last solar activity cycles (Kane, McTiernan, Hurley, 2005). Giant solar flares seem to have more than $10^{34}$ ergs already in the >20 keV electrons, and, importantly, comparable or larger energies may be present in other forms (kinetic energies etc.). The frequency distribution of flares as a function of their X-ray energy are well defined over a broad energy range from $10^{27}$ to $10^{33}$ ergs (Kucera et al., 1997). These energy values do not contain the optical and mechanical energies that are many times indicated as larger, sometimes much larger. Nevertheless, as a cautious value, we can take that flare theories should explain the source of flare energy at least up to $E_f \sim 10^{33}$-$10^{34}$ ergs, with $t_{imp} \sim 10^3$ s.

Now let us obtain a simple illuminating estimation of magnetic energies available for a big solar flare. With the magnetic field strength of an umbra $B_{umbra} \sim$ 3000 G, its linear scale $L_{umbra} \sim 10^4$ km, its length $L_{length} \sim 10^5$ km, the magnetic energy $E_B \sim (B^2/8\pi)V_{tube}$ contained in a volume of a flux tube connecting a big sunspot pair $V_{tube} \sim 10^{27}$ cm$^3$ will be $E_B$(flux tube)$\sim 10^{34}$ ergs. Now it is a matter of fact that the fields usually appear to be fairly potential, so there isn't a big fraction of free energy. Many people think they cannot relax to a completely potential state in any case because of inductive time scales. Therefore, allowing a factor $\eta < 0.1$ for the fraction of magnetic free energy, one big sunspot pair, in case of a complete transformation of all its free energy content into the flare site within the short time of the flare, the available flare energy is still $E_f < \eta*E_B$(flux tube)$\sim 10^{33}$ ergs. This simple estimation already shows that there is a problem with the magnetic explanation of flare energies supplied from the flux tube. Certainly, the total magnetic energy of a complete flux tube cannot be liberated in a flare also because observations do no indicate a characteristic disappearance of sunspots and their flux tubes after the onset of the flares. Now let us consider this point a bit more in detail.

Metcalf, Leka and Mickey (2005) had shown that there was an unusually large amount of free magnetic energy in NOAA AR 10486: $E_B$(available)$\sim(5.7\pm1.9)*10^{33}$ ergs. This value involves the free magnetic energy of the whole active region. Moreover, as Kane, McTiernan and Hurley (2005) pointed out, even when the total available energy in the active region is comparable to the energy released during the flare, release of all that energy during the short duration of the flare is expected to affect substantially the magnetic field structure of the active region, and, apparently, there are no observations indicating large scale changes in an active region after a large flare. Therefore it seems that the active region does not offer enough magnetic energy to supply the largest flares. The magnetic field outside of active regions is so weak that it does not represent an effective energy source. Certainly, even the assumed inflows could not transport most of the flare energy from another active region. Therefore, if the magnetic field of a whole active region would prove to be insufficient to supply the flares, than models working with inflow transport of magnetic field would fail.

A closer look to the magnetic free energy content of active regions reveals a still more fundamental problem for the exclusively magnetic flare theories. The current sheets seem to be produced after the flare onset as a quickly elongated feature extending from the loop top. Sui and Holman (2003) found observational indications that a current sheet formed between the top of the flare loops and the coronal source moved outward at $\sim$300 km s$^{-1}$, showing an upward expansion of the current sheet during the early impulsive phase. Sui et al. (2005) noted that the large scale current sheet formed due to the fact that the coronal source moved outward at a speed of $\sim$300 km s$^{-1}$, while the loop top moved at only $\sim$10 km s$^{-1}$, thus, the current sheet must have continuously elongated. We find these observations as indicative that the current sheet is produced in the flare process itself. This observation seems to indicate it is the mechanism that produces the current sheet and the related phenomena that is the flare driver, and the current sheet is not the cause but an important consequence of the flare. This proposal seems to be underpinned by some other recent observations.

Schrijver et al. (2005) realized that in the active region 10486 it is the emergence of currents into preexisting active region field configurations that appears to be required to drive flaring. They noted that the free energy of active region fields available for flaring is not built up by persistent stressing of the surface field, but instead emerges with the field from below the photosphere. This result is underpinned by the recent results of Wheatland and Metcalf (2006) who were able to determine the free energy of the whole active region 10486 two hours before the flare as $(2.6 \pm 0.11)*10^{33}$ ergs. Metcalf, Leka and Mickey (2005) determined that just after the impulsive phase the free energy of this active region is $(5.7\pm1.9)*10^{33}$ ergs. They remarked that this suggests that the free energy was increasing prior or during the flare, but the errors in 2005 were overly large to allow such a statement to be verified. Nevertheless, their recent measurements decreased the errors and now their suggestion is underpinned with a significant probability. In the case if such results will be more definitely established, not only the mechanism producing the flare but the process producing $\sim3*10^{33}$ ergs surplus in the free energy content of the active region magnetic fields within two hours before and during the flare should be also explained. The same situation is found by Zhang (2001, Table 1): the magnetic flux of the flaring active region begins to increase more than an hour before the flare. Similarly, with the help of the THEMIS telescope and the Michelson Doppler Imager (MDI) on SOHO space probe, Meunier and Kosovichev (2003) presented observations showing that at the time of the flare, a sharp increase of the positive magnetic flux by $10^{21}$ Mx occurred during at least 1 hour accompanied by strong flows both up and down especially at the time of the flare. Wang et al. (2004) also observed a rapid increase of magnetic flux in one polarity at the time of the flare, corresponding to a sudden emergence of new magnetic flux $10^{20}$ Mx hour$^{-1}$ at the site of the flare. Wang (2005) pointed out that recent BBSO and MDI magnetograph observations demonstrate more and more evidences of rapid changes of photospheric magnetic fields associated with the core regions of flares and CMEs. Metcalf, Leka and Mickey (2005) refer to Uchida and Shibata (1988) as a reconnection scenario that predicted energy increase of the active region during the flare. Actually, Uchida and Shibata (1988) suggested that the supply of energy to the loop top comes from the chromosphere or transition region immediately before the flare in the form of relaxing fronts of magnetic twist of opposite signs traveling within the twisted flare loop itself. Therefore, their model does not explain the transport of energy into the active region from outside. In contrast, our proposal is able to explain both the production of this enormous free energy surplus and the flare with one simple factor: the fast emergence of flux tube from below the photosphere, in accordance with the findings of e.g. Schrijver et al. (2005).

Now let us underpin this proposal by a relevant argument and by an estimation. In the standard flare models it is a hypothetical 'catastrophic loss of equilibrium' that is responsible for the triggering of the flare (e.g. Priest & Forbes, 2002). Recently, Lin (2005) suggested that the coronal accumulation of energy is built up by photospheric footpoint motions and the catastrophic loss of equilibrium is elicited by emerging flux tubes. Nevertheless, Schrijver et al.

(2005) demonstrated that shear flows related to coronal free energy require appropriately complex and dynamic flux emergence within the preceding ~30 hr and so they do not by themselves drive enhanced flaring. If photospheric footpoint motions are not able to support enough energy on the observed short period from the photosphere to the corona, another mechanism has to transport the photospheric energy into the corona, and this mechanism is directly related to the emerging flare loop itself. Now it is well known that flare loops are accelerated to speeds around 10 km s$^{-1}$ for the period of flare onset (e.g. Bruzek, 1964; Svestka, 1968; Tsuneta, 1993; Tsuneta, 1997; Kundu, 2001). The speed of the flare tube ~10 km s$^{-1}$ is remarkable especially in the light that the non-flaring emerging loops have much lower velocities. Fisher et al. (2000) reported on an average rise velocity of (non-flare) emerging flux tubes in the convective zone is around 0.01 km s$^{-1}$, corresponding to rise times through the convection zone 2-4 months. Zwaan (1992) noted that observations show loop tops passing through the photosphere with a speed of rise estimated at ~ 3 km s$^{-1}$. Schrijver et al. (1999) observed flux emergence at *chromospheric* heights with speeds averaged for a half an hour period as ~10 km s$^{-1}$. Caligari, Moreno-Insertis and Schüssler (1995) calculated that at 13,000 km below the surface the radial velocity of the flux loop's summit is ~ 0.5 km s$^{-1}$. This theoretical result is confirmed by observations when Kosovichev, Duvall and Scherrer (2000) with the help of time-distance inversion methods determined that the speed of the emerging flux tubes in the *subphotospheric* regions are around 0.5 to 1.3 km s$^{-1}$. By numerical experiments of the emergence of magnetic flux, Archontis et al. (2004) determined that the main body of the rising flux tube acquires a rise speed of about 1.7 km s$^{-1}$ while arriving at the *photosphere* yet still in the convective zone. By the analysis of observations, Spadaro et al. (2004) found that in the chromosphere the rising loop tops is characterized by velocities ~ 9 km s$^{-1}$. With the help of multi-height magnetic and velocity field measurements Choudhary, Suematsu and Ichimoto (2003) found that in the *photosphere* (as shown by Si I Dopplergram) the line-of-sight velocities characteristic of emerging flux is in the range 1.2 to -0.99 km s$^{-1}$, while in the chromosphere (as shown by He I Dopplergram) the velocities found are up to 5 km s$^{-1}$. Nagata et al. (2006) pointed out by high-resolution G-band observations that *photospheric* flux tubes move with a velocity 0.2-1.0 km s$^{-1}$. They pointed out that Stokes V profiles show flows with velocities up to 5 km s$^{-1}$ inside and outside the flux tubes. It is also clear that the emerging flux tubes rise in an environment in which the gas pressure drops rapidly with height, therefore the loop expands favorably towards its top. The expansion of loops is highly pronounced in the chromosphere and the corona. Therefore, it is not easy to subtract the effect of expansion when someone aims to determine the rise speed of the main body of the rising flux tube. Yet we can summarize the above shown observational and theoretical results finding that they consistently show that the main body of an average, non-flaring flux tube ascends upwards in the subphotospheric regions with speeds ~0.5 to ~2 km s$^{-1}$.

The rise speed of the main body of the flaring loops around photospheric heights, in contrast, is much higher. Already Bruzek (1964, its Fig. 6) noted that there is a close association between loop prominences (flaring loops) and flares, and found that such flare-associated loop prominences rise with a speed ~10 km s$^{-1}$ from below the photosphere. Svestka (1968) realized that most of the observed slowly ascending limb flares may be explained by loop prominences ascending with speeds around 8 km s$^{-1}$. The point is that most of the limb and disk flare loops are "slowly ascending" with very similar speeds, therefore, the emergence of flare loops with 8-10 km s$^{-1}$ may explain most of the flares. Tsuneta et al. (1992) observed a limb flare with the Yohkoh X-ray telescope and found that the loop overlying above the flare loop starts to ascend at the flare onset with a speed of 10-30 km s$^{-1}$, accompanied by a footpoint separation rate of a similar speed (a value larger by an order of magnitude of the footpoint separation speed of non-flaring emerging flux tubes). Tsuneta (1993) described the evolution of the flare in terms of flux tubes and their rise speeds. He found that the rapidly expanding flare loop of the 1991 December 2 flare appeared 5 minutes before the flare from below the photosphere, and its speed around flare onset is ~10 km s$^{-1}$, a value decreasing to 5 km s$^{-1}$ after 30 minutes from flare maximum. Tsuneta (1997) in his Figure 3 (upper panel) presented observations of the evolution of flare loop heights below and during the flare. These results tell that the flare loop had shown a quasi-constant rise speed from below the flare onset until ~30 minutes after the flare maximum; from that time onwards its rise speed is decelerated. Kundu et al. (2001) observed that the main flaring loop of the 1993 November 11 event had emerged from below the photosphere just below the flare onset and started to rise with a speed about 8-10 km s$^{-1}$. We think these observations demonstrate that flaring loops emerge from below the photosphere with a velocity 8-10 km s$^{-1}$ that is much larger than the photospheric rise speed of quiescent, non-flaring emerging flux tubes.

It is important to observe that flaring loops seem to be accelerated not only by magnetic buoyancy, since non-flaring loops have similar fields but lower rise speeds. The existence of a non-magnetic acceleration factor seems to require the acceleration of material within and below the flux tubes. The average density in the subphotospheric regions is higher than $4*10^{-7}$ g cm$^{-3}$. These flux tubes has a volume around V~$10^{27-28}$ cm$^3$. Now the acceleration of material extending to a similar volume to such high speeds requires significant energies $E_{acc}>4*10^{32-33}$ ergs. This energy is in the same range required to fuel the magnetic free energy of flaring active regions by the recent results of Metcalf et al. (2005) and Wheatland & Metcalf (2006). Their results, in agreement with many others (Ishii, Kurokawa and Takeuchi 1998, 2002, 2004, Kurokawa, Wang and Ishii 2002, Schrijver et al. 2005) involve that the free energy of the active region have to be supplied by emerging flux tubes. Now the most crucial emerging flux tube is just the flare loop. It is this context that offers a far-reaching context to our results that bubbles rise to the subphotosphere just by speeds ~10 km s$^{-1}$, and their energies are in the range up to ~$10^{34}$ ergs and more. Now if exclusively magnetic flare theories cannot explain by a coronal 'catastrophic loss of equilibrium' the huge energies present in emerging flare loops and their acceleration to high speed for the time of flare onset, they fail to explain the most energetic aspect of flare phenomena.

There are many observations showing that flares are driven by emerging flux tubes. For example, Green et al. (2003) found that the majority of CMEs and flares occur during or after new flux emergence. Ishii, Kurokawa and Takeuchi (1998, 2000, 2004) as well as Kurokawa, Wang and Ishii (2002) emphasized that "the emergence of twisted flux bundles is the energy source of strong flares". The same conjecture is drawn by Schrijver et al. (2005). Again, our argument works: if the flare is to be explained by an explosive coronal process liberating an enormous amount of radiative energy in a relatively short time and in a small volume; and if the energy source of the flare is the emergence of flux tubes from below the photosphere, than the only possibility is that the explosive coronal process is the result of the emerging flux tubes and not the other way around. Other observations underpin this argument. Seely et al. (1994) found that the energy is deposited in a small volume at the top of the flaring loop often as small (or smaller) than a single pixel (1,800 km*1,800 km). Zhang (2001) determined that the source of the Moreton wave in the X12/3B flare was originated between the photosphere and the upper chromosphere. Yamaguchi et al. (2003) found that the Moreton wave of the 1991 June 4 flare was emitted from a flare bright point that initially showed the form of a loop. We can identify the source of the Moreton wave with the looptop or the nearby region above it. Martin (2004) had shown that the X-class flares had a bright core in the chromosphere and the energy spread from this source. Movie 8 of Schrijver et al. (1999) shows the time evolution of the limb flare 1998 May 19. A slide show view of this movie demonstrates that the bright regions (associated with relatively dense and excited material) propagate consequently from below upwards before flare onset as well as during and after. Immediately before the flare onset, at 07:57:46, a bright vertical feature extends from the looptop of the flaring loop upwards. This bright vertical feature developed in relation to a feature ejected from the top of the flaring loop at 07:53:16 in the form of an ascending bright knot.

The above described observations show that the bulk energy of flares is supplied not from the free magnetic energy of the coronal region but in the form of currents and their hosting flux tubes emerging from below the photosphere. In the light of the above listed theoretical and observational results, we became aware of a fundamental problem: If the bulk energy of the flare is related to the emerging flux tubes, and if these flux tubes represent energy in the (sub)photospheric and chromospheric regions, how will be the energy of the flare localized into the looptop and the region above it? In other words: there is a missing link, a process concentrating the (sub)photospheric/chromospheric energy related to the flare loop into the coronal flare site. Fortunately, there are many important observations indicating the nature of this connection, like the ones we already mentioned above by Sui and Holman (2003) and Sui et al. (2005) indicating that the site of the primary energy release (the current sheet) develops from the looptop as an elongated vertical feature determined by the rise of the flare looptop with ~10 km s$^{-1}$ and the rise of the overlying loop with ~300 km s$^{-1}$. These observations indicate that the site of the primary energy release of the flare (the current sheet) is enveloped into antiparallel field lines that are elongated with ~300 km s$^{-1}$. In this context, the fundamental question of flare origin is: what is the mechanism driving the generation of this vertical magnetic feature?

If the energy of the flare is not supplied from the corona, one has to show which energy source can suffice, and provide also quantitative evidences. The detailed calculations that we realized considering the origin and development of the hot bubbles offer a new solution for this problem, nicely fitting to the findings of e.g. Schrijver et al. (2005). Table 4 shows that hot bubbles, when arrive to the subphotospheric regions, have tremendous energies, sufficing the energy requirement. The main obstacle is removed, since the basic problem of flare energy source can be solved within the frames offered by our model calculations, and so the main problem of the solar flares is resolved. Certainly, in order to make our promising result usable, a whole list of secondary problems should be solved. One of these accompanying problems is that we observe flares not in the subphotospheric, but in the coronal regions, and, apparently, with smaller masses. Let us present here some simple considerations that may relate our findings with more detailed observations.

The missing link is supplied by the observation that in the subphotospheric regions the hot bubbles are continuously accelerated in an environment in which the sound speed is smaller and smaller outwards. Our calculations indicate that hot bubbles may easily reach the threshold of sound speed where they may suffer sonic boom. If a hot bubble suffers a sonic boom, it will be destroyed by the developing 'compression wall' at the front of the generated shock wave (offering as a side result, a clue to the generation of flare related shock waves). Depending on local conditions, like the structure and strength of the magnetic fields, the material of the abruptly destroyed hot bubble compressed into the shock front will be transformed into a particle beam directed upwards, towards the loop top of the flux bundles that are carried by the hot bubble itself from below.

The point is that an energetic particle beam traveling between subphotospheric regions and loop tops with a velocity of 100 to 1000 km s$^{-1}$ is hardly visible until it interacts with the decelerating influence of relatively strong magnetic fields at the loop top. This is a crucial point in relating our numerical results to flare observations – and there are some promising answers offered. Already Kleczek (1964) pointed out, that energy in the form of corpuscular radiation is practically invisible in the photosphere and in the chromosphere. He referred to Warwick (1962) who argued that a 300 MeV particle beam may reach the photosphere from below a 300-500 km depth and we could not observe it in the optical region. Orrall (1964) commented this point, that "if the particles are injected into the loops at 1000 km s$^{-1}$, we should not expect to see these in the corona - that is, in the coronograph - against the sky. We should not expect to see them at all." Actually, the generation of particle beam injected towards the corona from below the photosphere should at first lift off the material of the photosphere. As a result, below the flare sites we will observe practically subphotospheric regions, i.e. somewhat hotter regions. Apparently, this is the phenomenon that is already observed by Machado and Linsky (1975) who remarked that "most of the flares show signs of photospheric heating during the flares. The energy of the photospheric heating is comparable to that of the chromospheric and coronal ones. This suggests that the energy source of solar flares is at least subphotospheric."

Energy considerations may turn to be fruitful when applied at the level of the photosphere and at coronal heights. Since we found that a particle beam may transport a large part of the energy content of the hot bubble from below the photosphere to the coronal loop top, the same amount of energy, $E_f \sim 10^{33}$ -$10^{34}$ ergs, is present in the photosphere as well as in the corona.

What is the size that is needed for a region with the temperature of photospheric regions, $T \sim 10^4$ K, to represent an energy content of e.g. $E_f \sim 10^{34}$ ergs? The answer of our model is simple. From the relation $E_f = cmT$, we obtain $m_b = E_f/cT \sim 5*10^{21}$ g. With photospheric densities, $\rho \sim 4*10^{-7}$ g cm$^{-3}$, we obtain for the linear size of the region $L \sim 10^9$ cm, similar to the size of the flare kernels. On the other hand, since at the local sound speed the thermal and kinetic energies are equal, a region with similar size and having a velocity of the sound speed has a similar kinetic energy, $E_{kin} \sim 10^{34}$ ergs. Now we have a suitable mechanism to concentrate this energy into the shock front, compressing the hot bubble's material and generating the particle beam having much higher velocity: the sudden destruction of the high energy hot bubble, allegedly by the sonic boom (and/or by magnetic reconnection below the photosphere). Let us apply, as a first step, a most simple approach. Since, as we saw, a speed of v~10 km s$^{-1}$ corresponds to T~10$^4$ K, and the energy depends on the square of v, while depends linearly on T, 100 km s$^{-1}$ for the particles corresponds to 10$^6$ K kinetic temperature at the flare site and 1 000 km s$^{-1}$ to 10$^8$ K. Actually, the site of primary energy release cannot be regarded as thermalized. Nevertheless, with some restrictions, one can speak of the temperature of the HXR source around the flare onset, and its value is found frequently in between 10$^6$ and 10$^8$ K; in energy units, 0.13 to 13 keV. Of course, nonthermal electrons corresponding to higher but unthermalized 'temperatures' represent a significant part of the energy of the flare. Therefore, on observational grounds we may require that the particle beam should correspond to a velocity $v_{particles}$>100-1 000 km s$^{-1}$. By energy conservation we obtain that the mass carried by the particle beam $m_{particles}$ will be reduced in comparison to the mass of the subphotospheric hot bubble in a ratio $R_{mass}$~((10 km s$^{-1}$)/$v_{particles}$)$^2$<10$^{-2}$-10$^{-4}$, offering a mass for the particle beam $m_{particles}$~ $m_b$*$R_{mass}$<5*10$^{17}$-5*10$^{19}$ g for the largest flares. Since the size of the primary energy release has a spatial scale L~10$^4$ km, a mass of a large flare with $E_f$~10$^{32}$ ergs will correspond to $m_{particles}$~5*10$^{15}$ g, a value that would increase the density at the loop tops to n~1.8*10$^{11}$ cm$^{-3}$. Again, this is consistent with observations. Already Zirin (1988, 409) noted that "it is really hard to understand how density can peak at the loop tops" in defiance to hydrostatic equilibrium. Actually, the density in the loop top may increase by an order of magnitude at flare onset in a duration less than a minute. Recently, Veronig, Brown and Bone (2005) noted that plasma density is already enhanced at the flare onset, and increases for a peak density of n~(1.3-2.2)*10$^{11}$ cm$^{-3}$. In a looptop having a linear size ~10$^4$ km, such a density enhancement corresponds to a mass enhancement of ~5*10$^{14}$ g.

We mention that recent theories and observations indicate the presence of subphotopsheric explosive processes in relation to solar eruptions (e.g. Hiremath, 2005b; Kosovichev, Duvall 2005). Some authors assumed that the subphotospheric events are related to subphotospheric magnetic reconnection. Not excluding this explanation, we propose also to consider that these events may be related to the dynamics of high-speed emerging flux tubes and the generation of particle beams. Kosovichev and Duvall (2003) already observed that the growth of active regions is characterized by multiple emergence of magnetic flux structures propagating *very rapidly* in the upper convection zone. In the context of our findings, the application of acoustic tomography to reveal the presence of high-speed subphotospheric bubbles before the flare beneath flare sites could serve with further observational evidences. Unfortunately, the time resolution of such measurements is at present very low, around 8 hours (Kosovichev and Duvall, 2004).

One of the remaining basic problems, as we indicated above, is how the X-type neutral point is formed as a result of the upward motion of the flare loop (Tsuneta, 1997). In our model, it is natural to consider that it is the until now 'missing element', the energetic particle beam that elongates the field lines upwards, and so it generates the hard X-ray source at the looptop at the flare onset as well as the vertically elongating magnetic structure, in a way that it becomes elongated above the looptop afterwards due to particle beam injection from below. The expected result is an antiparallel, elongated magnetic semi-island growing upwards from the loop top, as a 'neck' with a 'head' which develops at the frontside into a separated 'plasmoid'. Now since the particle beam will not only push the field lines upwards, but at the same time a large part of the particle flux will follow the field lines bending towards the chromosphere, it will generate chromospheric evaporation, injecting also particle beams back into the loop top region from the chromospheric footpoints. These returning beams when transferring their momentum to field lines around the looptop will generate the elongated X-formed magnetic structures on the two sides ('arms'). We suggest particle beam injections from below as the mechanisms by which the observed characteristic X-shaped configuration as well as the antiparallel field line structure, i.e. the 'current sheet' develops (Fig. 7; from Fig. 5a of Tsuneta, 1993; see also Fig. 8).

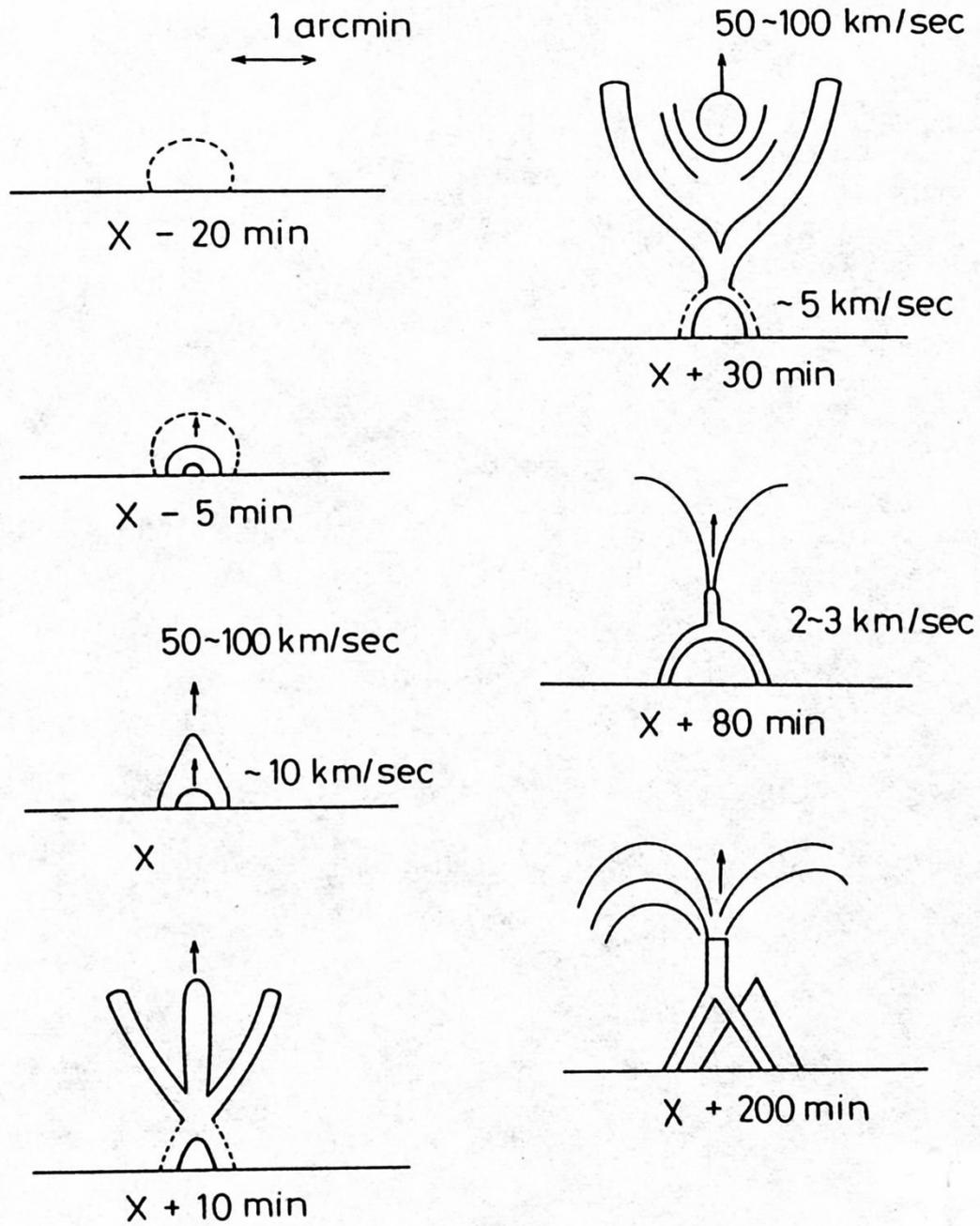

Fig. 7; from Fig. 5a of Tsuneta, 1993, shows the time development of the magnetic configuration before and after the limb flare.

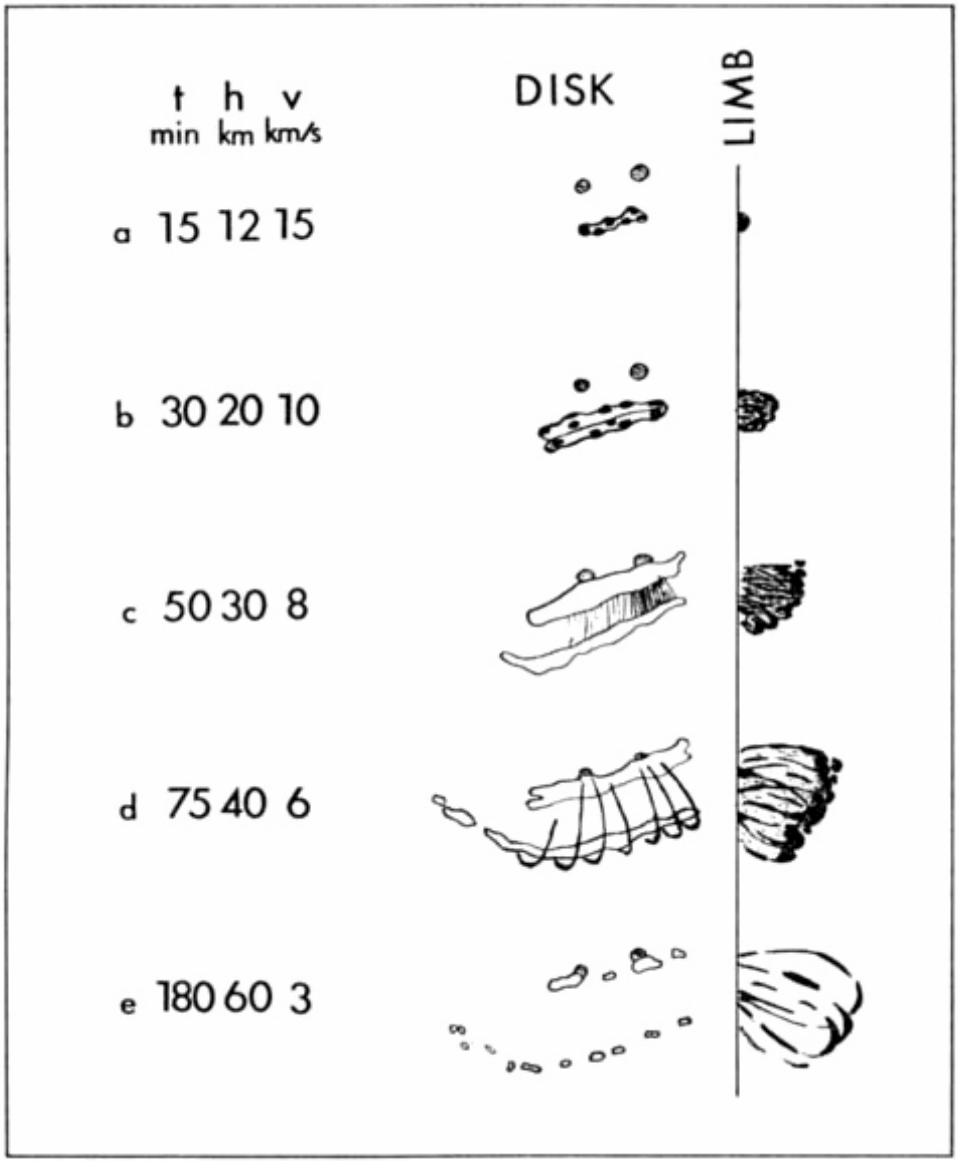

Fig. 8, from Bruzek, 1964, its Fig. 6, shows the time development of the magnetic configuration before and after the flare on the disk and at the limb.

The proposal presented here telling that particle beam injected from below into the top of the flare loop offers solutions to some problems, namely:
i.) Our mechanism explains the basic problem how to generate the magnetic topology necessary for reconnection;
ii.) Our mechanism explains why the loop tops show upward elongated cusp-like structures instead of downward concave intrusions generated by particle beams injected from higher lying regions downward, as it is indicated by simulations (Forbes, Malherbe, Priest 1989, Fig. 2; Forbes, Malherbe, 1991, Fig. 5). Present-day pictures on the mechanisms generating the flares (e.g. Fig. 1 of Lin & Soon, 2004) assume the presence of reconnection outflow injected into the loop top from above, but ignores the observable consequences that would be concave structures instead of the observed upward elongating antiparallel semi-islands and cusp-like structures;
iii.) Our mechanism explains why the primary energy liberation occurs mostly above the loop tops but below the X-point. Saint-Hilaire and Benz (2002) had shown that most of the initial energy first appears as energetic electrons in the lower, stationary part. Since the particle beams are injected in our model from below, the X-point is generated at the loop top and is driven upwards by the particle beam/plasmoid later on; therefore the primary energy release is produced in our model in between the loop-top and the X-point.

Our model has observational consequences also for the chemical composition of the flare material. Apparently, in the temperature range we explored here, nuclear heating does not help significantly the bubbles to reach the convective zone. As our results show, the reason is that the timescale of nuclear heating below $T \sim 1.5*10^8$ K is larger than the timescale of cooling of the bubble, $\tau_{cool} < \tau_{nucl}$. Our tentative calculations indicated that thermonuclear runaway really might develop above $T > 1.5*10^8$ K, where $\tau_{nucl}$ becomes lower than $\tau_{cool}$. In that case, one could expect flare-related chemical anomalies with larger amplitude. Actually, the flare-related chemical anomalies represent a long-standing unsolved issue of solar physics (see e.g. Kerridge, 1989; Sterling et al., 1993, Waljeski et al., 1994; Fludra and Schmelz, 1999, Feldman, Landi, and Laming, 2005). For example, Waljeski et al. (1994) presented results from measurements of soft X-rays (SXR) line and broadband intensities. They showed that for the observed active region the absolute abundances of the low first-ionization-potential (FIP) elements (Fe, Mg) are enhanced in the corona relative to the photosphere by a factor of 6 to 31, in a way that the abundances of the high FIP elements (e.g. Ne, O) are also enhanced by a factor larger than 1.75. Flare seed material plays a significant role in iron-rich gradual events. Recent measurements have shown that the heavy ion composition shows distinct differences from solar wind material, enhancements that are primarily due to the properties of the seed population of the flares (Mason et al., 2005). Although this question is still poorly understood, it seems possible to proceed by the development of new methods measuring the amount of absolute abundances of heavy elements in the flare. Our results predict that such absolute enhancements are actually present.

There are strong indications that the flare-related chemical anomalies represent not relative, but absolute enhancements (Grandpierre, 1996). For example, the nitrogen enigma (Kerridge, 1989) states that the $^{15}N/^{14}N$ rate is enhanced by 50%, from a value $2.9*10^{-3}$ of $3*10^9$ years ago to a present day value of $4.4*10^{-3}$ (Kerridge et al., 1991). It is just the opposite change of what the stellar evolution models predict. To produce the observed enhancement not only in the solar wind but also in the convective zone as a whole, would mean that the rate of this enhancement is so enormous that it exceeds the values by many orders of magnitude allowed by the standard models for the solar convective zone. This circumstance suggests that the solar surface is connected to the core by channels which are isolated from the convective zone, connecting the central regions with the subphotospheric regions directly. These central regions have to produce significant amount of heavy elements, like e. g. $^{15}N$, which is possible only above $10^8$ - $10^9$ K, i. e. in a local explosive process. Our model can be tested independently when the chemical anomalies related to solar flares will allow determining absolute chemical abundances of the different elements present in the flare site. We add that recent solar models (Basu and Antia, 2004; Bahcall and Pinsonneault, 2004; Turck-Chieze et al., 2004; Guzik, Watson and Cox, 2005) pointed out that solar models evolved with standard opacities and diffusion treatment give poor agreement with helioseismic inferences for sound-speed and density profile, convection-zone helium abundance, and convection-zone depth. Varying the input parameters, none of the variations tried completely restores the good agreement attained using the earlier abundances. The problem is so severe that Guzik, Watson and Cox (2005) recommended considering accretion of material depleted in the more volatile elements C, N, O, Ne and Ar. Our theory offers a natural alternative, again, without any further assumptions, by the occasional heavy element enhancements of some hot bubbles and related flare materials.

We mention that although the two crucial elements offered by our calculations for flare models, the high velocity hot bubbles in the subphotosphere and the particle beam injected from below the photosphere into the loop tops, are new propositions, they have the merit to draw together most of the results of competing flare models into one coherent picture. Namely, the newly arisen picture is consistent and complementary with the classical Hirayama (1974) model, with the Heyvaerts et al. (1977) emerging flux tube model, with the global picture of Priest (1995). It offers an energetically more suitable explanation for the generation of efficient acceleration of non-thermal particles (Hudson and Khan, 1996) and is consistent with the current-centered flare model (Melrose, 1997) with the idea that the particle beam interrupts the current at the loop top; it fits with the plasmoid model (Ohyama and Shibata, 2002) since the plasmoid is the result of the interaction of the upward injected particle beam with the looptop. The model presented here suggests the presence of a primary nonthermal energy source in relation to preflare velocity fluctuations (Nigro, Malara and Veltri, 2005) generated by the shock waves and particle beams since the sonic boom occurs before the flare and below the photosphere.

The results obtained here call attention to the principal possibility that local metainstability of generation of heated bubbles may explain the rigid rotation of some activity centers (Spence et al., 1993), as well as the existence of sunspot nests (Castenmiller et al., 1986, De Toma et al., 2000), hot spots (Bai et al., 1995), and active longitudes (Bai et al. 1995, Bai 2002, 2003). We note that the appearance of hot bubbles in the solar core may provide certain dynamism to the solar radiative interior and "very slow" mixing. The dynamic nature of the solar core (Grandpierre, 1990, 1996, 1999) is indicated not only by the lithium problem (Deliyannis et al., 1998, Zahn, 2001) and related problems with a need of a kind of mixing in the radiative interior, but also by the anomalously slow rotation of the core.

Now that we saw some successful explanation of a whole list of basic and yet unsolved problems, let us summarize here some predictions of our model that can be tested by future observations:
i.) hot bubble(s) rising in the subphotosphere beneath the flare site accelerated to sonic speeds ~8-10 km s$^{-1}$;
ii.) shock waves are generated in the subphotosphere;
iii.) particle beam is generated in the subphotosphere and is injected upwards;
iv.) the material of the photosphere is lifted up before the particle beam, therefore the surface of the Sun is hotter underneath of the flare site than elsewhere;
v.) flare material is enhanced in heavy elements not only as a result of selective electromagnetic processes;
vi.) reconnection topology is generated as a consequence of the process driving the coronal primary energy release of the flare, formed by particle beams injected from below;
vii.) measurement of the rise speed of the main body of flare loops;

viii.) the solar core have a dynamic nature that can be tested by e.g. detecting g-mode solar oscillations (Turck-Chieze et al., 2004).

Obtaining detailed data like local correlation tracking (LCT) based on speckle masking white-light images, near-infrared (NIR) continuum images at 1.56 mum, and G-band images could be helpful to test some of these points. Such data about flare-related photospheric flows are indicated to be a critical observational diagnostic for the evolution of magnetic fields in solar active regions (Deng et al., 2004).

Let us call attention to some new contexts that are shown as of interest in our understanding solar flares. For example, it is important to obtain more data on the time evolution of emerging flux tubes before and during flares (like Fig. 6 of Bruzek, 1964 and Fig. 5a of Tsuneta, 1993), also in dependence of depth/height beneath/above the photosphere. Movies of limb flares like that of Schrijver et al. (1999) could be especially helpful (see Figs. 9-22).

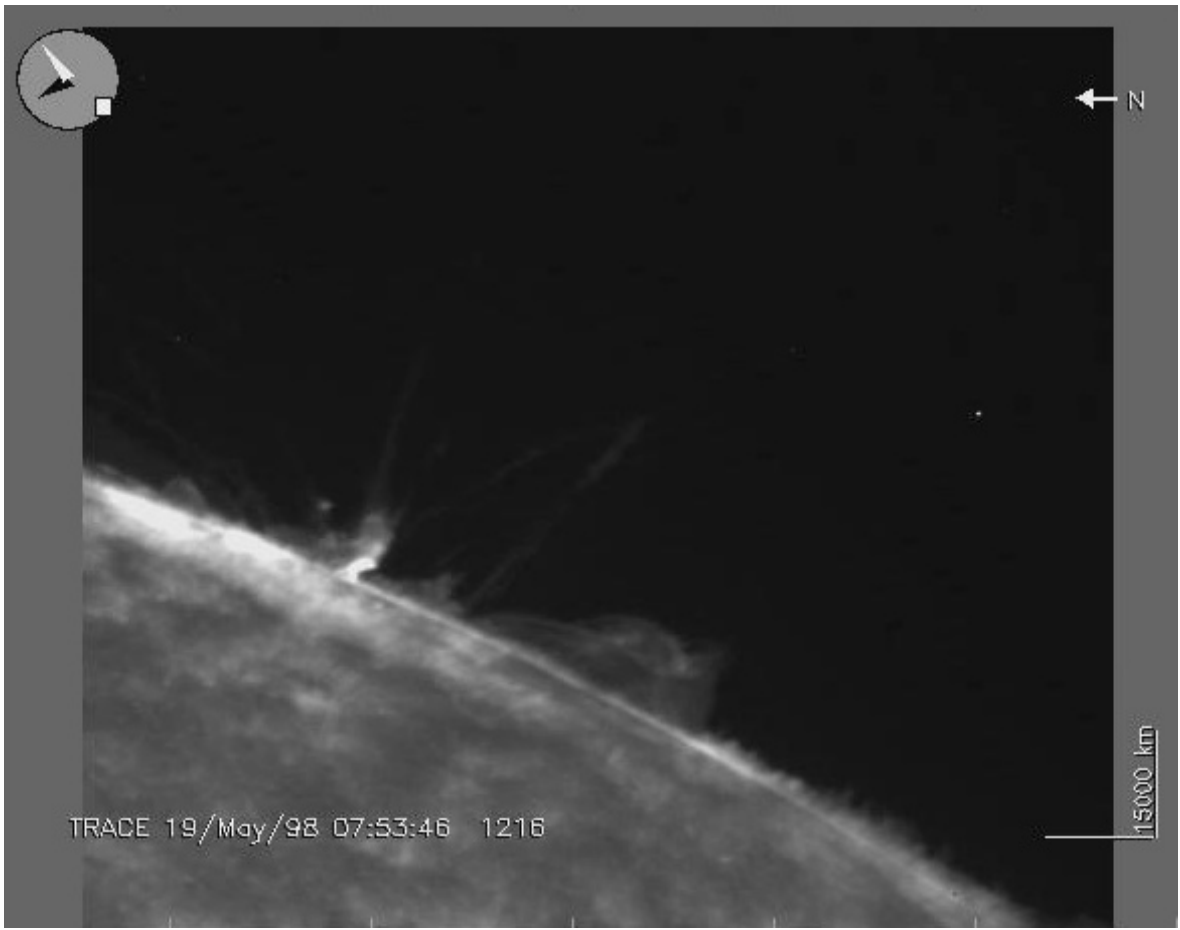

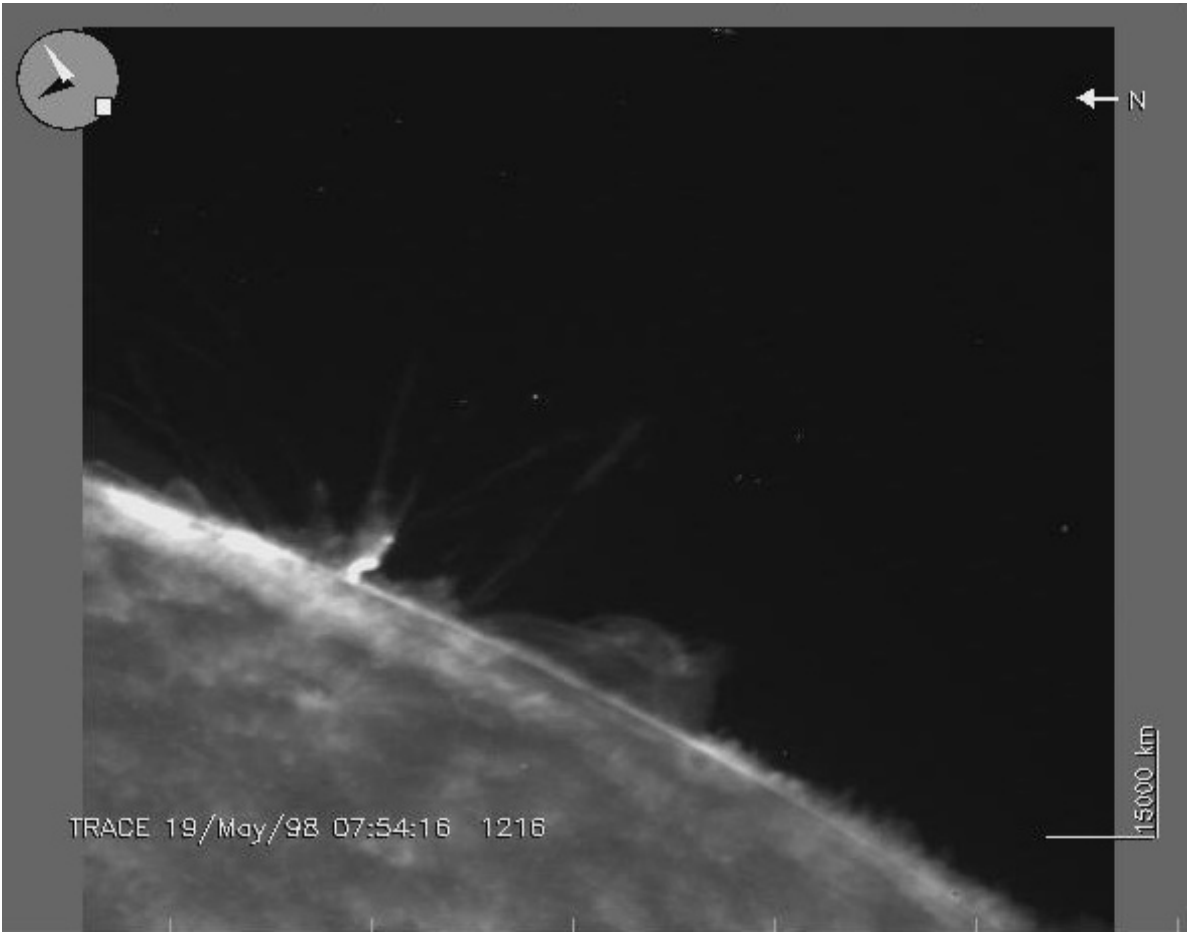

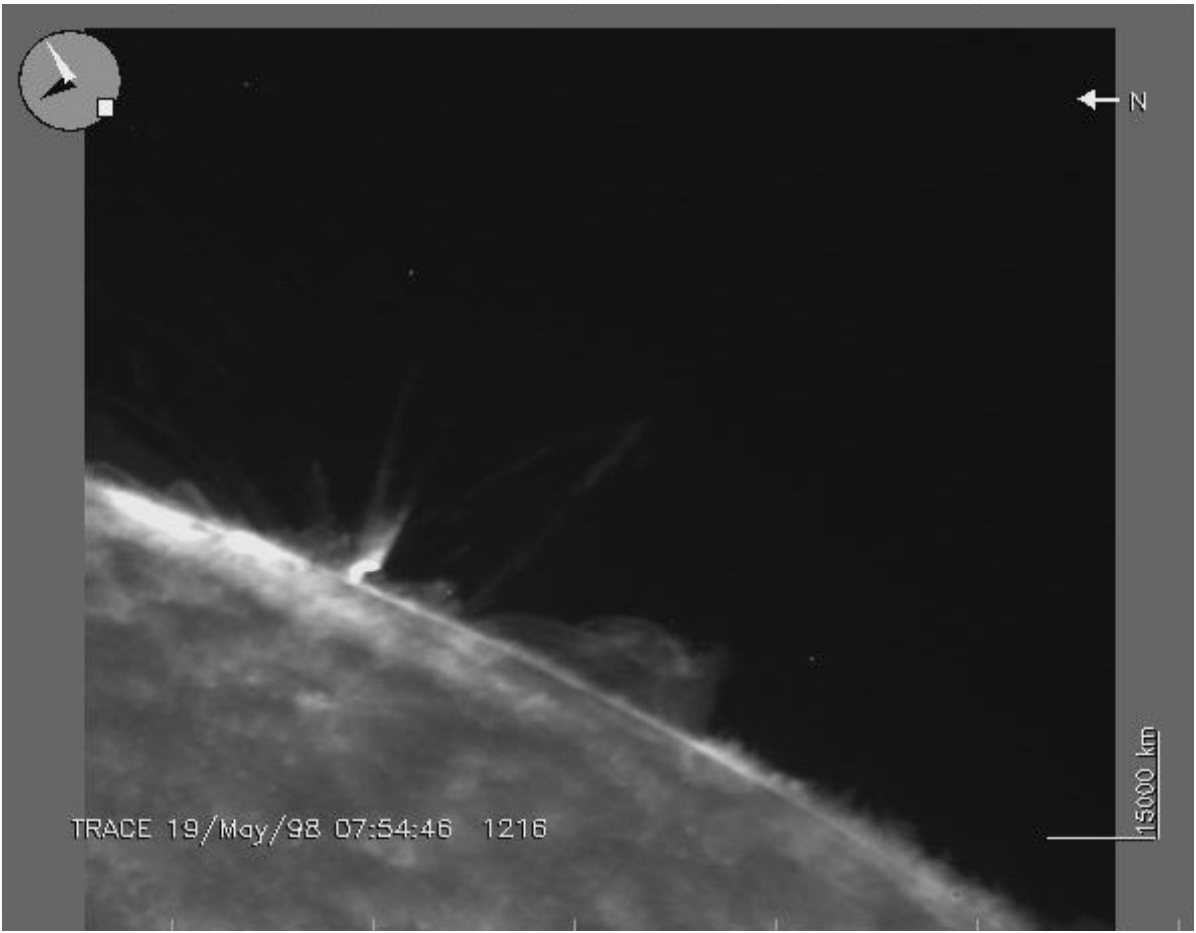

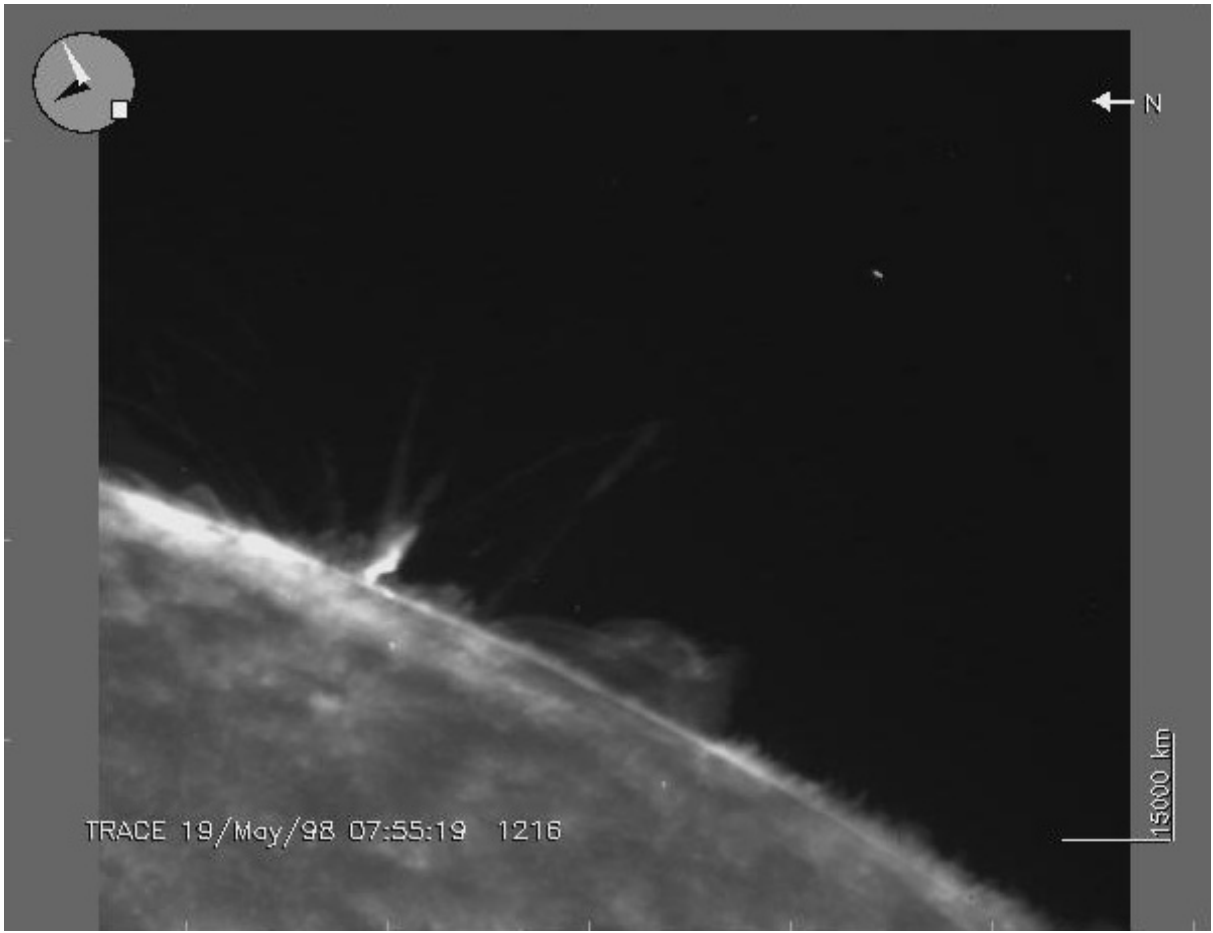

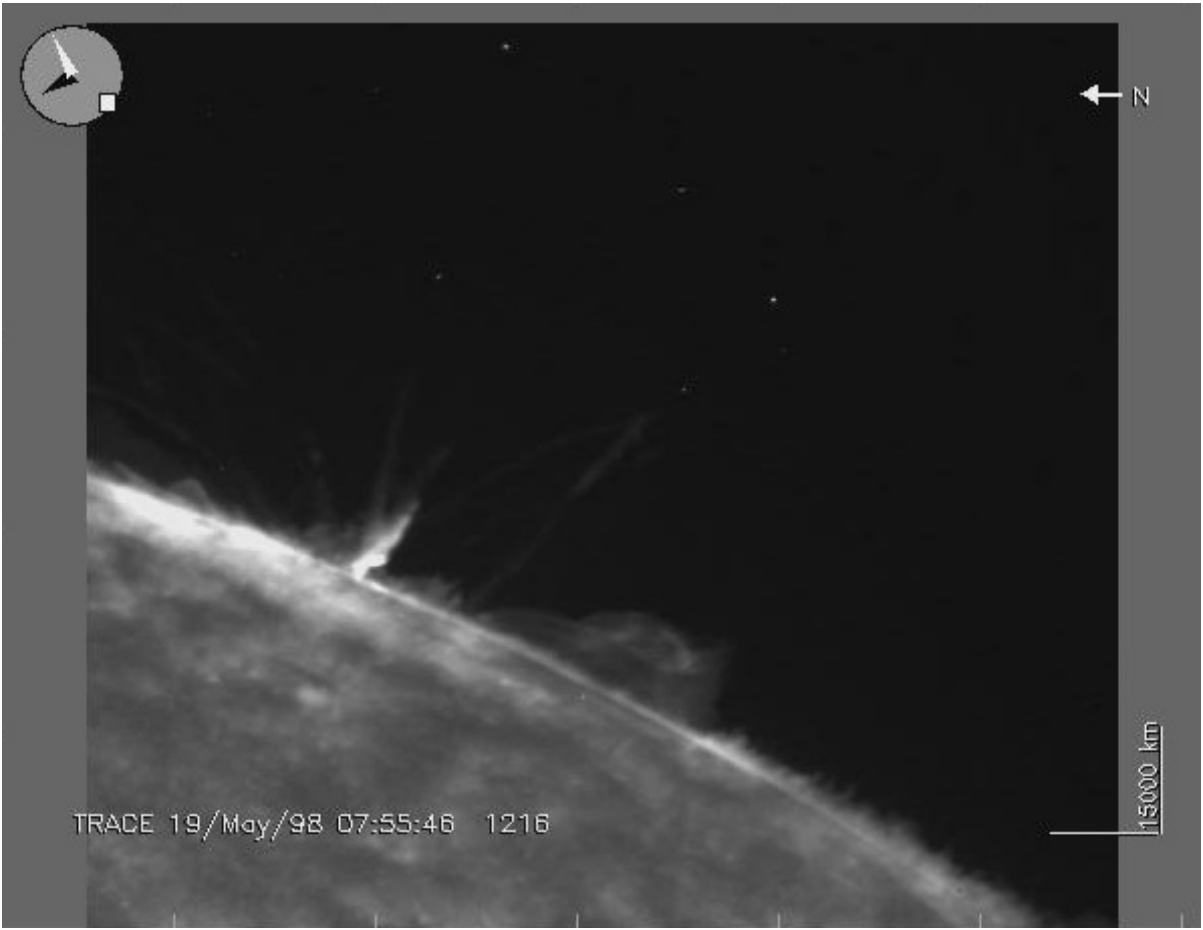

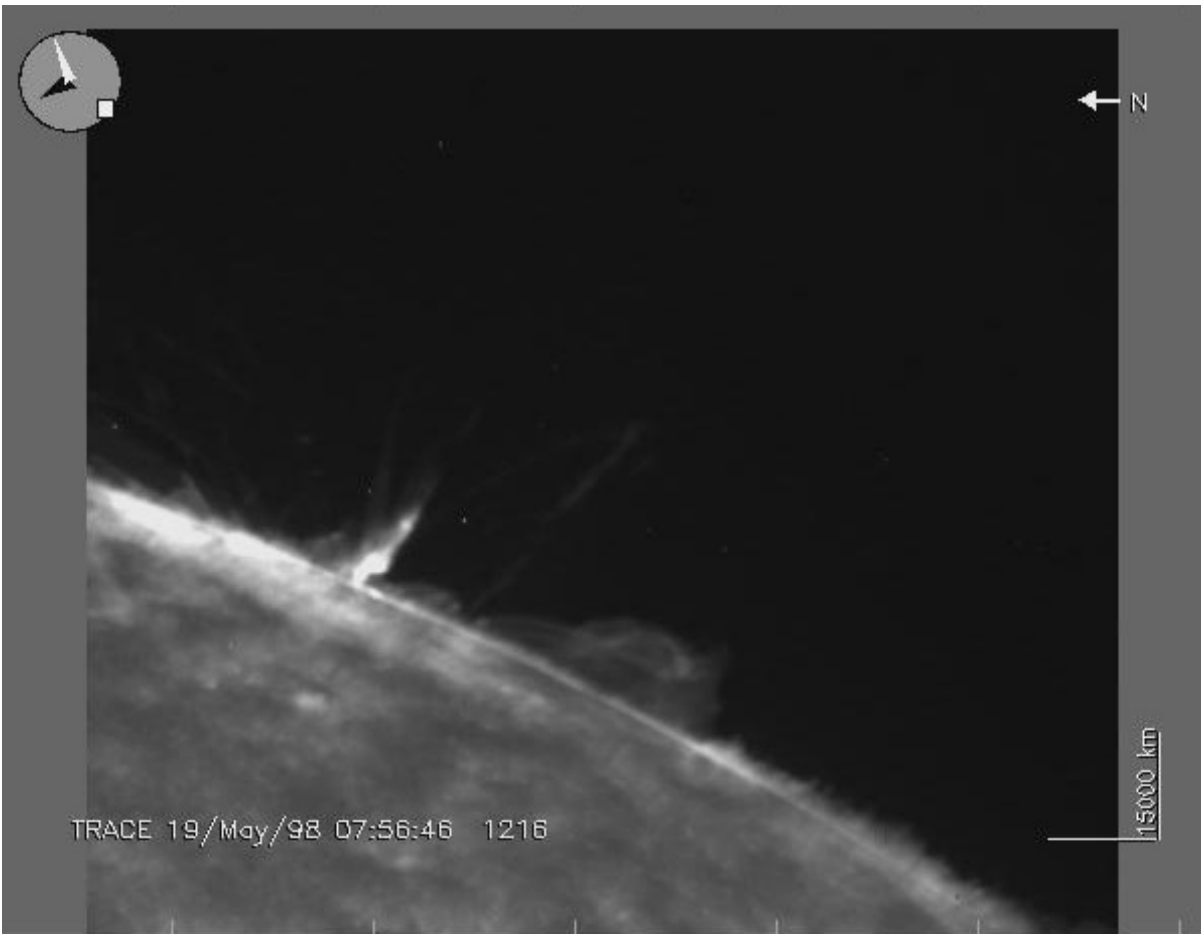

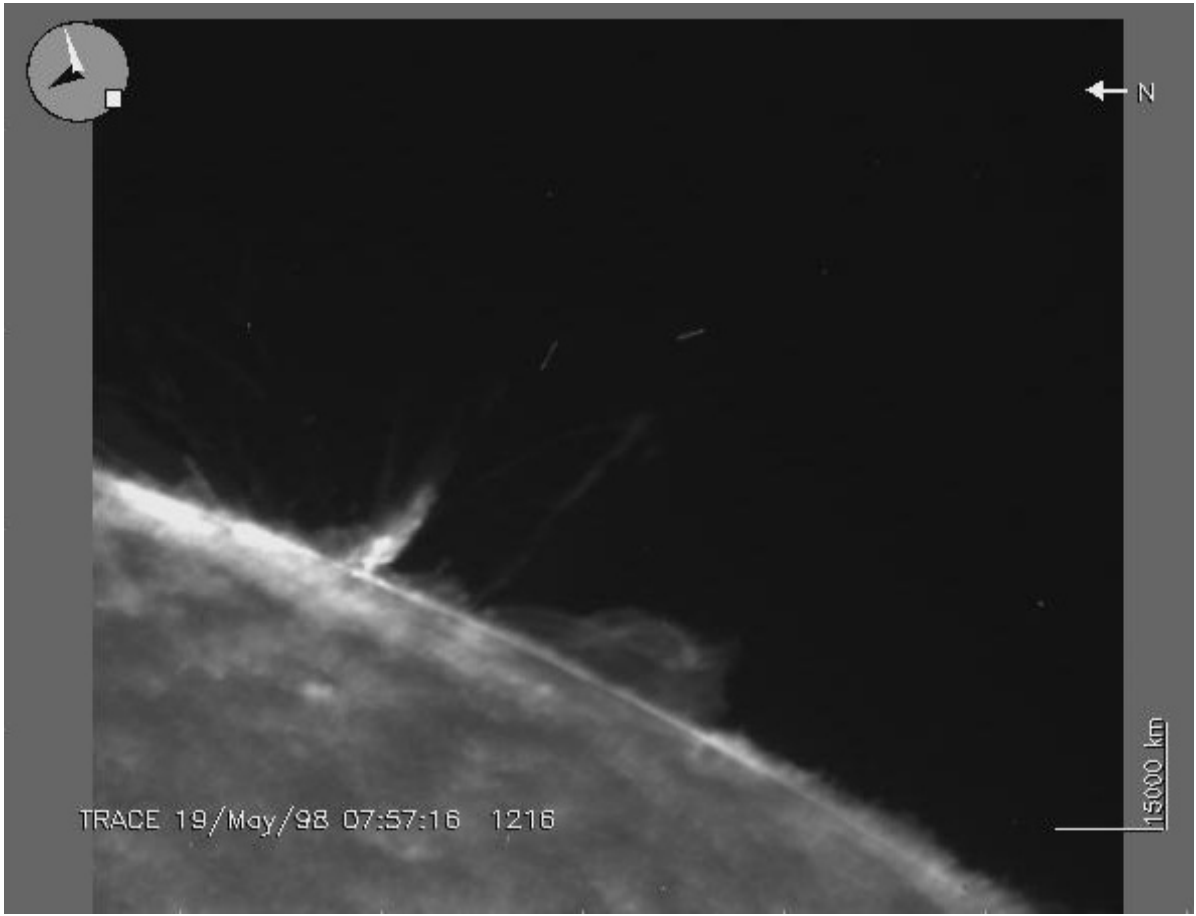

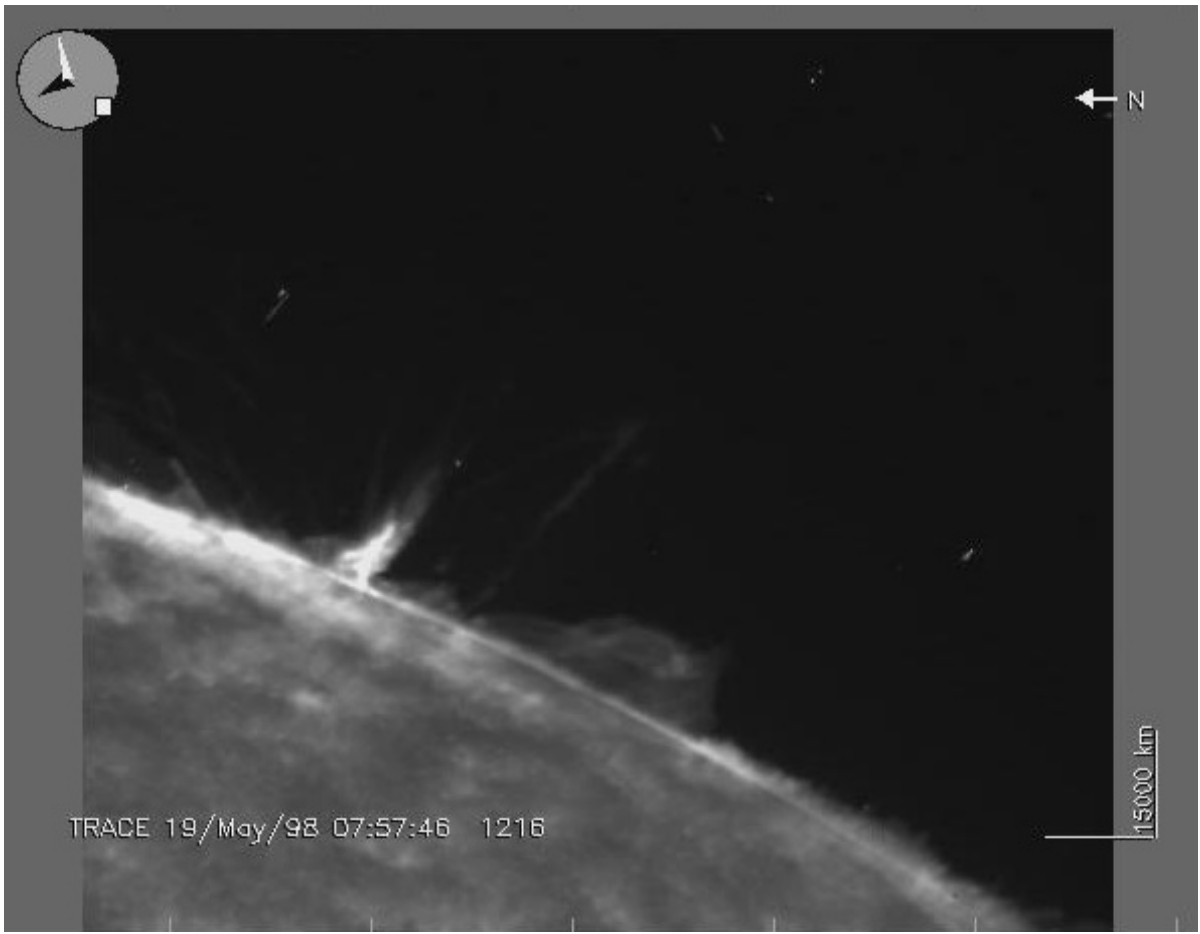

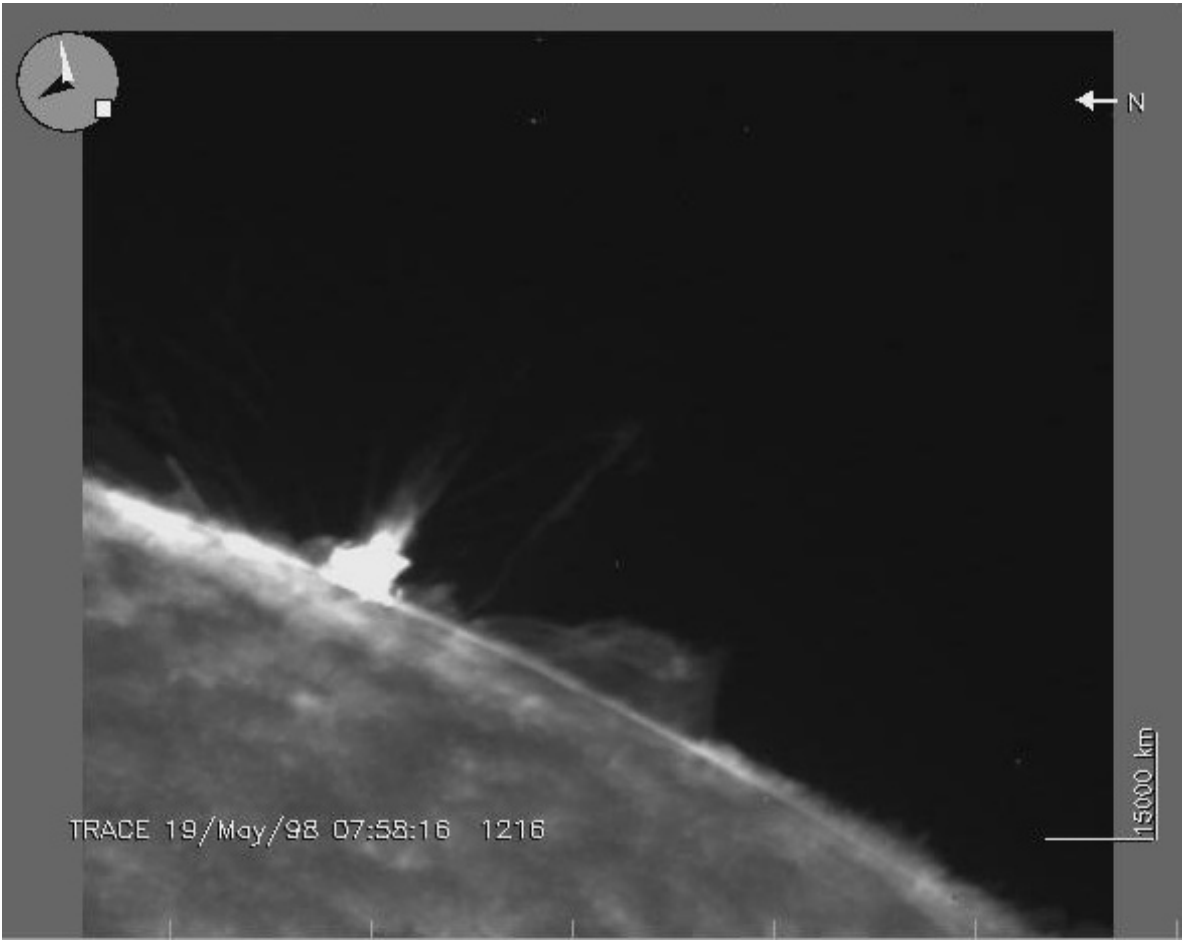

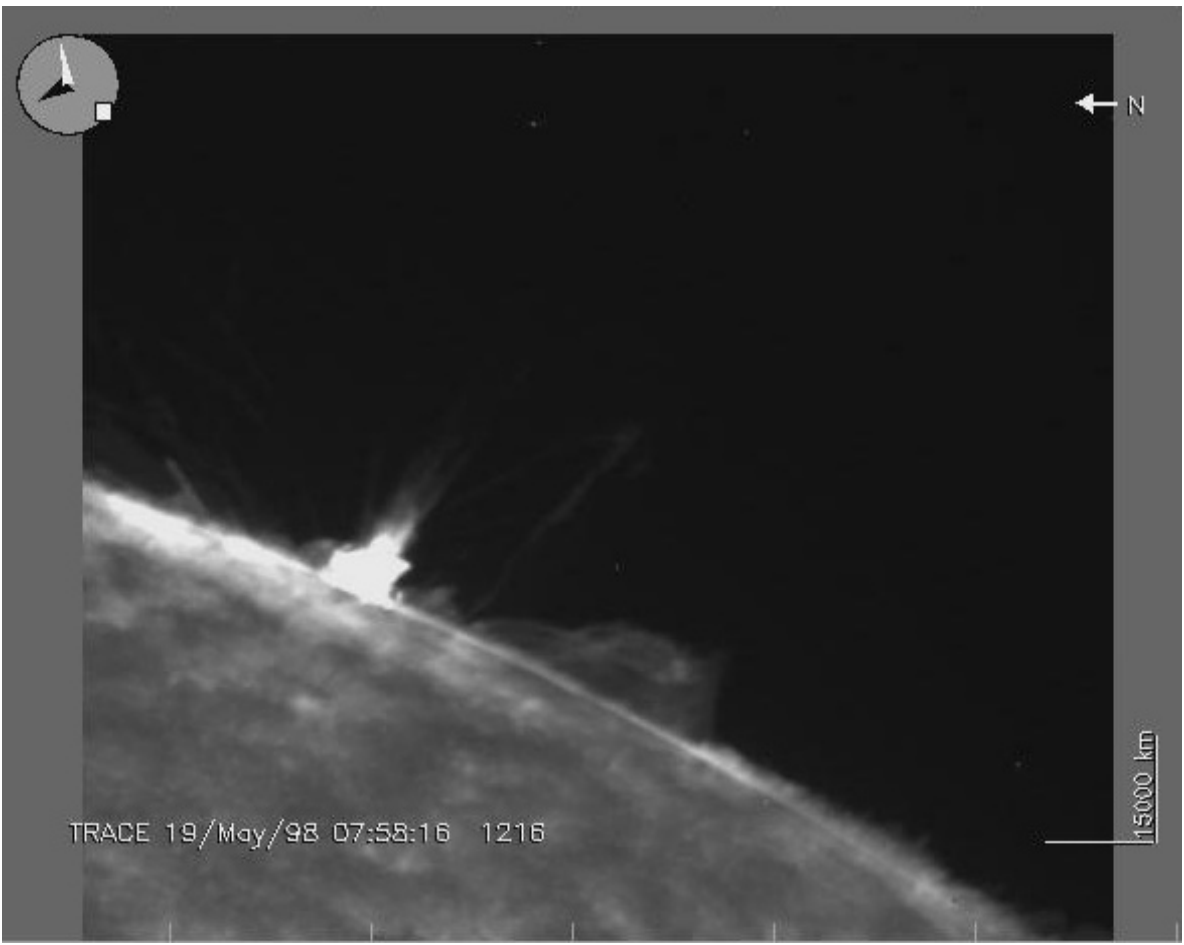

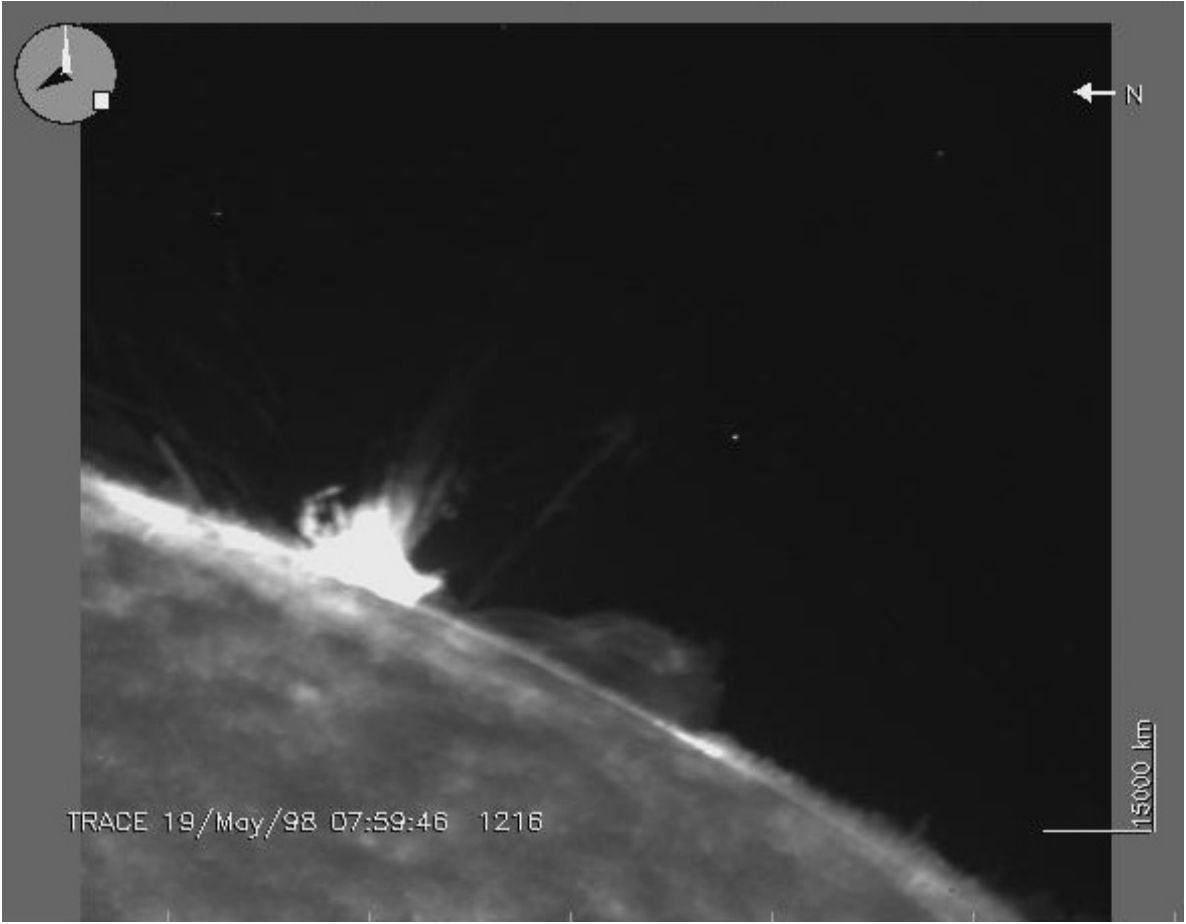

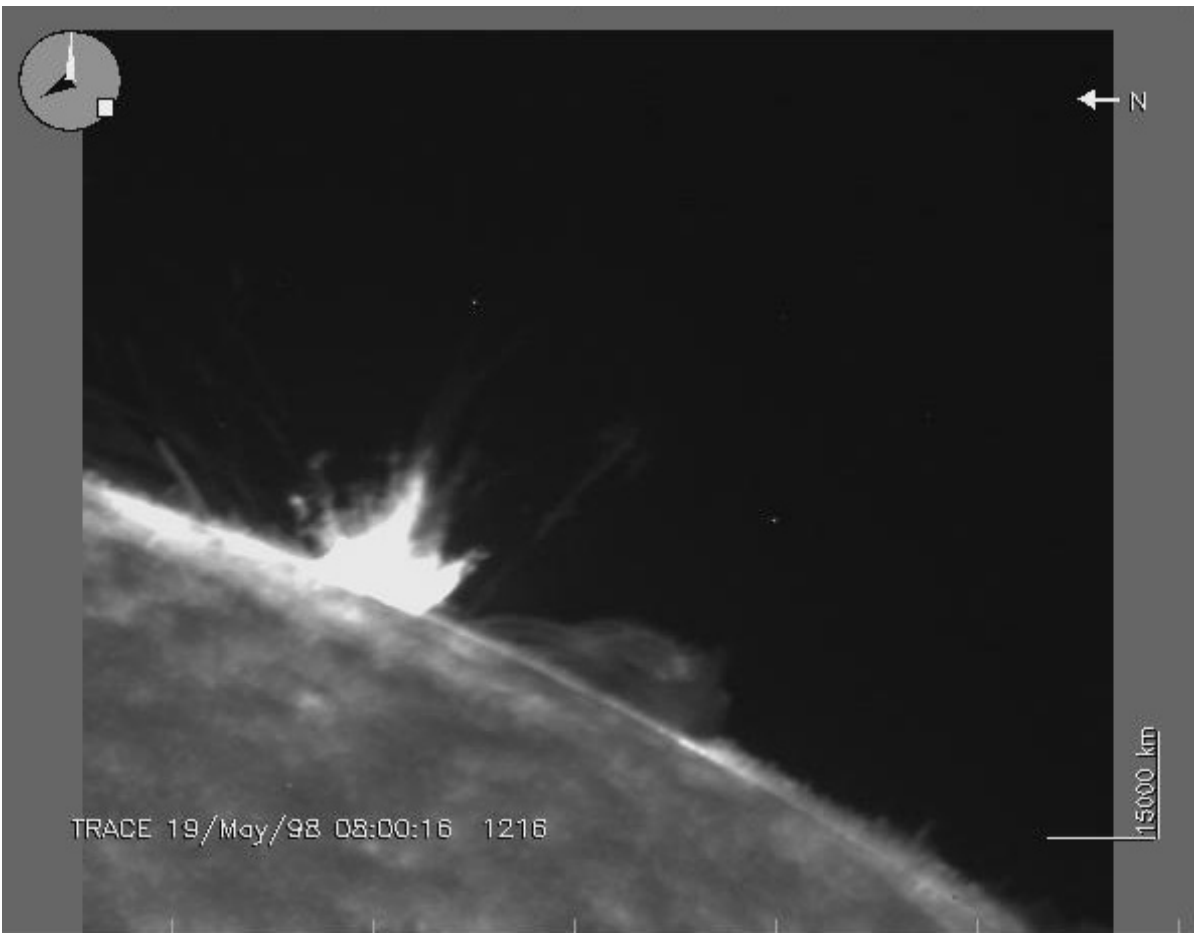

Figs. 9-20 are from Movie 8 of Schrijver et al. (1999). It shows the time evolution of the limb flare 1998 May 19.

Our model is the first to solve some unsolved basic problems of solar activity. At the same time it indicates new problems, points out to new contexts, necessitates new observations and theoretical efforts, therefore it seems to be fruitful enough to consider it as worth to attention. In the face of the presented evidences, the model obtained here is indicated to be plausible. It answers some of the long-standing problems of solar physics and has predictions for observations that are planned to be realized in the near future. Revealing the presence of metastabilities in the solar core may help our understanding of the different types of instabilities, angular momentum dissipation, spin-down of the solar core and the dynamism arising from its plasma nature. The metainstabilities calculated in this paper may be directly relevant in our understanding of the generation of the solar, and, in general, stellar activity cycles.

## 7. ACKNOWLEDGEMENTS

It is a pleasure to express our gratitude to Profs. Hudson, Priest, Forbes, Sturrock, Jakimiec, and Kálmán, for the useful notes that were helpful in the preparation of the final version. The author wishes to express his thanks to Dr. Hiremath for the valuable comments on the earlier version of the manuscript, and to his friend, Jean Drew for her galvanizing inspirations and lecturing the English.